# A revisit of the density gradient theory and the mean field theory for the vapor-liquid interface system

Hongqin Liu (刘洪勤)

Integrated High Performance Computing Branch,

Shared Services Canada, Montreal, Canada

Emails: hongqin.liu@ssc-spc.gc.ca; hqliu2000@gmail.com

## Abstract

In this work we define a mean-field crossover generated by the Maxwell construction as the dividing interface for the vapor-liquid interface area. A highly accurate density-profile equation is thus derived, which is physically favorable and leads to reliable predictions of interfacial properties. By using the density gradient theory and a mean-field equation of sate for the Lennard-Jones fluid, we are able to extensively explore the interface system in terms of the Gibbs free energy, the Helmholtz free energy and heat capacity. The results show that the mean-field dividing interface is the natural extension of the Widom line into the coexistence region. Hence the entire phase space is coherently divided into liquid-like and gas-like regions in all three (temperature-pressure-volume) planes. Some unconventional behaviors are observed for the intrinsic heat capacity, being positive in low temperature region while negative in high temperature region. Finally, a complete picture of the mean-field equation of state is unfolded: all three solutions to a vapor-liquid equilibrium problem have their respective significances.

## Introduction

Theoretical understanding and accurate description of a vapor-liquid interface system are important for practical applications such as modeling surface tension and adsorption. Here an interface system is composed of homogeneous bulk vapor and liquid phases and an inhomogeneous interfacial region (interface area). Classic equilibrium thermodynamics and statistical mechanics are applicable to the stable bulk phases. In particular, in the framework of the mean-field theory (equation of state, EoS), saturated volumes are obtained from the two physical solutions to a vapor-liquid equilibrium (VLE) problem subjected to the Maxwell construction. In dealing with the interface area, several theories have been developed for describing the heterogeneous feature. The density gradient theory founded by van der Waals [1] and further developed by Cahn and Hilliard [2] has been widely employed in the area. The density functional theory has also found applications to the interface systems [3]. Statistical mechanics theory, in particular, the correlation function method provides a tool to calculate the local density and related properties [1,3,4,5]. For thorough reviews of the researches in this area, the reader is referred to the references [1,3-7].

In the interfacial region, discontinuous bulk densities for the vapor and liquid phases are bridged together with a position-dependent local density, $\rho(z)$, known as the density profile, where $z$ is the position variable normal to the interface (or surface). Computer simulations are used to provide the density profile data at different conditions [8-11]. As the density profile is known, various properties can be expressed in terms of the local density. A closely related concept is the Gibbs dividing interface, which divides the interface area into two regions. The definition of the dividing interface is somewhat arbitrarily [1,3]. Conventionally, the arithmetic mean of vapor and liquid densities is used to define the diving interface [1,3]. Namely, the origin of the interface area ($z = z_0$) is defined at the average density. This definition leads to zero-excess (or deficit) molecules on the interface, which is required by a phenomenological definition of surface tension [1] and hence accepted, unnecessarily, in various density profile expressions. However, there is an unphysical aspect with this interface: the decay length (or thickness) of the vapor side is the same as that of the liquid side. Physical observation tells that the decay length of vapor boundary is less than that of liquid boundary [6]. In addition, for the truncated and shifted Lennard-Jones fluid, Stephan et al. [12] show that the density gradient theory and the density function theory based on the algebraic mean interface fail in predicting the local maximum difference between the normal and tangential components of the pressure tensor reported by computer simulations.

The two-phase coexistence region diminishes as temperature rises and finally disappears at the critical point. The system above the critical point is known as the supercritical fluid. Traditionally, the supercritical fluid is considered to be a uniform phase. However, in last decades or so, some outstanding features have been discovered in the supercritical region. In particular, a line defined with the loci of the local-maximum heat capacity, $C_p$, known as the Widom line [13], divides a supercritical area into gas-like and liquid-like regions [14-22]. Nevertheless, the relationship between the behaviors in the supercritical region and those in the interface area is seldom addressed. Currently, the Widom line is considered as a smooth continuation of the saturated pressure in the pressure-temperature space, while in the pressure-volume (density) and the temperature-density spaces, the extension of the Widom line becomes bifurcated.

In this work, we revisit the theories for the vapor-liquid interface system, namely the density gradient theory [1-3] for the heterogeneous interfacial contribution and the mean-field theory for the homogeneous contribution. The objective is threefold: (1) defining a new dividing interface and thus proposing a novel density profile model for the interfacial region; (2) extensively exploring the behaviors of the vapor-liquid interface system with the density gradient theory and the mean-field theory by using the new density profile model; (3) addressing the relationship between the behaviors in the supercritical region and those in the interfacial region.

The paper is organized as the following. In the next section, we provide some theoretical background required for descriptions of the vapor-liquid interface system, along with a new density profile expression. To illustrate the applications of the density gradient theory incorporated with the mean field theory, we make use of the Lennard-Jones fluid as a sample system and calculation procedures and results will be presented and discussed in the third section. The conclusions and further discussions are provided in the last section. The derivation of the density profile expression is given in Appendix A. Finally, more calculation details for the Lennard-Jones fluid and supplementary results are provided in Appendix B.

## Theoretical background for the vapor-liquid interface system

In this section we present some theoretical background and major results at macroscopic (phenomenological) and microscopic (position-dependent) levels for succeeding calculations and analysis. In this work, a planner interface perpendicular to the surface with one coordinate, $z$, is considered. We start with the grand potential ($\Omega$) for an interface system [3]:

$$\Omega = -PV + \gamma\mathcal{A} \tag{1}$$

where $P$ is the pressure, $V$, the total volume (the lower case, $v = 1/\rho$, refers to molar volume), $\mathcal{A}$, the interfacial area and $\gamma$, the surface tension. The above definition is independent of the position of the Gibbs dividing interface defined at $z = z_0$ [1,3]. Another important property is the Gibbs adsorption (of component *i*) which can be derived from the well-known Gibbs adsorption equation:

$$s^{(s)}dT + \sum_i \Gamma_i d\,\mu_i + d\gamma = 0 \tag{2}$$

where $s^{(s)}$ is the excess entropy of the surface per unit area, $\Gamma_i$, the adsorption of component *i* on the surface, $\mu_i$, the chemical potential (the Gibbs free energy for one component, or pure, system). At constant temperature, for a one-component (pure) system, Eq.(2) reduces to:

$$d\gamma = -\Gamma dG \tag{3}$$

The quantities in Eq.(2) and Eq.(3) are defined for a stable uniform fluid without involving any dividing interface. On the other hand, the surface tension can be calculated from the position-dependent pressure difference:

$$\gamma = \int_{-\infty}^{\infty} \left(\frac{d\gamma}{dz}\right) dz = \int_{-\infty}^{\infty} [p_N(z) - p_T(z)]\, dz \qquad (4)$$

where $p_N(z)$ and $p_T(z)$ are the normal (to the interface) and the tangential components of the pressure tensor, respectively. Eq.(4) presents the relation between a local property, $p_N(z) - p_T(z)$ and a macroscopic property. The adsorption can be calculated from the density profile, $\rho(z)$. For vapor phase, it reads:

$$\Gamma_v = \int_{-\infty}^{z_0} [\rho(z) - \rho_v]\, dz \qquad (5)$$

and for liquid phase:

$$\Gamma_L = \int_{z_0}^{\infty} [\rho(z) - \rho_L]\, dz \qquad (6)$$

where the subscripts, $v$ and $L$ refer to vapor and liquid phases, respectively. $\rho_v$ and $\rho_L$ are the saturated (equilibrium) bulk densities. Apparently, $\Gamma_v > 0$, namely the adsorption at the vapor side is an excess, and $\Gamma_L < 0$, the adsorption at the liquid side is a deficit.

In the conventional (classic) model, the dividing surface is defined such that

$$\int_{-\infty}^{z_0} [\rho(z) - \rho_v]\, dz + \int_{z_0}^{\infty} [\rho(z) - \rho_L]\, dz = 0 \qquad (7)$$

and the algebraic mean is adopted at $z = z_0$:

$$\rho(0) = \frac{\rho_v + \rho_L}{2} \qquad (8)$$

Before moving on, we need to make a few remarks on Eq.(7) and Eq.(8). Macroscopically, the excess free energy of an interface is given by [1]: $F^s = \gamma\boldsymbol{\mathcal{A}} + \boldsymbol{\mu n^s}$ where $\boldsymbol{\mu n^s}$ represents the sum of chemical potential multiplied by the excess surface density. If the surface tension is defined as $\gamma = F^s/\boldsymbol{\mathcal{A}}$, then the dividing interface should be chosen in such a way that $\boldsymbol{\mu n^s = 0}$ and this is where the equal-molar definition, Eq.(8), comes into play [1]. For this reason, historically Eq.(8) has been adopted for various density profile models. However, if the grand potential is used, Eq.(1), such a restriction is not required. Moreover, with the density gradient theory, as the free energy is expressed as Eq.(9) below, the restriction, Eq.(7), (hence Eq.(8)) is not required either [1]. By the way, it can be shown that the term $\sum_i \Gamma_i d\,\mu_i$ in Eq.(2) or $\Gamma dG$ in Eq.(3) is invariant with the position change of the dividing interface [1]. In summary, the choice of the interface is arbitrary in the context of the density gradient theory [1,3], which makes a different choice physically acceptable.

In the following we briefly summarize some major equations for the Helmholtz free energy (the total free energy), $\boldsymbol{\mathcal{F}}$, the local Gibbs free energy, $\mathbf{G}[\rho(z)]$ and related properties for the interface system. Full coverages can be found from Refs.[1-3]. According to the density gradient theory [2], for a pure system the total free energy can be obtained from the free energy density $\boldsymbol{f}(\rho)$ by integration over the entire volume of the system, $\mathcal{V}$:

$$\begin{aligned}\boldsymbol{\mathcal{F}} &= \mathcal{V}\int \boldsymbol{f}(\rho)\, dV \\ &= \mathcal{V}\int [f_0(\rho) + k_1\nabla^2\rho + k_2(\nabla\rho)^2 + \cdots]\, dV \qquad (9)\end{aligned}$$

where the coefficients, $k_1 = [\partial f/\partial\, \nabla^2\rho]_0$, $k_2 = [\partial^2 f/(\partial\langle\nabla\rho\rangle)^2]_0$, are the functions of the uniform density. In Eq.(9) the local (position-dependent) free energy density, $\boldsymbol{f}(\rho) = a(\rho)\rho$, is expanded about the free energy density with the uniform density, $f_0(\rho) = a_0(\rho)\rho$. All odd-order terms vanished since $\boldsymbol{f}(\rho)$ is a scalar and it must be invariant with respect to the direction of the gradient [2,3]. By applying the divergence theorem, one gets:

$$\begin{aligned}\int (k_1\nabla^2\rho) dV &= -\int (dk_1/d\rho)(\nabla\rho)^2 dV \\ &+ \oint (k_1\nabla\rho\cdot\boldsymbol{n})\, dS \qquad (10)\end{aligned}$$

where the last term accounts for the boundary exchange and the integration is over the entire surface of the system. For a system with no external field, we can choose a boundary in such a manner that $\nabla\rho\cdot\boldsymbol{n} = 0$. In addition, here we only consider the flat interface with one direction, $z$ and $\nabla = d/dz$ etc. Then from Eq.(9) and Eq.(10) we have the well-known total free energy expression:

$$\boldsymbol{\mathcal{F}} = \mathcal{A}\int_{-\infty}^{\infty} \left[f_0(\rho) + \frac{1}{2}m\left(\frac{d\rho}{dz}\right)^2\right] dz \qquad (11)$$

From the above treatment, Eq.(9), we see that Eq.(11) is effectively exact up to the 3rd order by omitting the 4th and higher order terms. In Eq.(11), the coefficient, $m$, is related to $k_1$ and $k_2$ as shown by Eq.(9) and Eq.10) and known as the influence parameter [1,3]. Generally speaking, the influence parameter is a function of the local density and can be evaluated from the direct correlation function by the following [3,6]:

$$m(\rho) = \frac{k_B T}{6} \int_0^\infty c(r,\rho) r^2 d\vec{r} \tag{12}$$

where $k_B$ is the Boltzmann constant and $c(r,\rho)$, the direct correlation function. For most applications, the density on the left hand side of Eq.(12) is treated as position-independent for a stable fluid. For a pure and stable fluid under normal pressure the density is uniquely dependent on temperature, $\rho(T)$, and therefore the influence parameter can also be considered as temperature-dependent, $m(T)$. In some cases, this parameter is even roughly treated as a constant [1,6].

For an interface system to reach to equilibrium, the total free energy, Eq.(11), is minimized, and one gets:

$$\frac{m}{2}\frac{d}{dz}\left[\left(\frac{d\rho}{dz}\right)^2\right] = \frac{d\Omega}{dz} \tag{13}$$

Upon integration:

$$\frac{m}{2}\left(\frac{d\rho}{dz}\right)^2 = \Delta\Omega = \Omega - \Omega_b = \Omega + P^s \tag{14}$$

where, $\mu(\rho) = df_0/d\rho$ and at equilibrium:

$$\Omega = \left[a_0\big(\rho(z)\big) - \mu^s\right]\rho(z) \tag{15}$$

$P^s$ and $\mu^s$ are the pressure and the chemical potential at equilibrium condition, respectively. The free energy $a_0\big(\rho(z)\big)$ can be evaluated by a mean-field EoS with $\rho(z)$ as the density. Then the surface tension can be calculated by the following:

$$\gamma = \sqrt{2m}\int_{\rho_v}^{\rho_L} \sqrt{\Delta\Omega}\, d\rho \tag{16}$$

From Eq.(14), we also have

$$\gamma = m\int_{-\infty}^{\infty}\left(\frac{d\rho}{dz}\right)^2 dz = 2\int_{-\infty}^{\infty}\Delta\Omega(\rho)\, dz \tag{17}$$

Eq.(16) and Eq.(17) can be used to evaluate the influence parameter from surface tension data. However, due to the approximations made in the density gradient theory, the values of the influence parameter estimated from Eq.(16) and Eq.(17) are not entirely consistent with each other. In most cases Eq.(16) is adopted to estimate the influence parameter. Another important result is obtained from the integration of Eq.(14):

$$z = z_0 + \int_{\rho_0}^{\rho}\sqrt{\frac{m}{2\Delta\Omega}}\, d\rho \tag{18}$$

As shown by Eq.(15), with an EoS at a given temperature, the expression of $\Delta\Omega$ is known, then Eq.(18) provides the density profile numerically if the influence parameter can be obtained elsewhere, such as from the direct correlation function, Eq.(12).

For our succeeding analysis, we present an important equation for the Gibbs free energy. The generic grand potential $\Omega_n$ and free energy $\mathcal{F}_n$ has the following relation [3]:

$$\Omega_n[n(z)] = \mathcal{F}_n[n(z)] - N\mathbf{G}[\rho(z)] \tag{19}$$

where $N$ is the total number of particles and $\mathbf{G}$ is the Gibbs free energy. For an interface system, both $\Omega_n$ and $\mathcal{F}_n$ are functional of the generic density profile, $n(z)$. As the system reaches to equilibrium, $n(z) \to \rho(z)$, $\Omega_n[n(z)] \to \Omega[\rho(r)]$, $\mathcal{F}_n[n(z)] \to \boldsymbol{\mathcal{F}}[\rho(z)]$, and $\Omega_n[n(z)]$ reaches to a minimum value [6], therefore Eq.(19) yields:

$$\left.\frac{\delta\Omega_n[n(z)]}{\delta n(z)}\right|_{\rho(z)} = \left.\frac{\delta\mathcal{F}_n[n(z)]}{\delta n(z)}\right|_{\rho(z)} - \mathbf{G}[\rho(z)] \tag{20}$$

At equilibrium, the derivative of the generic free energy in Eq.(20) is the same as that of $\mathcal{F}\big(\rho(z)\big)$ with respect to $\rho(z)$. Throughout this work, after Ref.[3], we define an intrinsic (local) property of the interface system as the sum of a mean-field homogeneous contribution and a heterogeneous counterpart (the density gradients), such as the intrinsic free energy given by the integrand of Eq.(11). From Eq.(20) and (11), we finally have the intrinsic Gibbs free energy:

$$\mathbf{G}[\rho(z)] = \mu\big(\rho(z)\big) - m\frac{d^2\rho}{dz^2} - \frac{1}{2}m'\left(\frac{d\rho}{dz}\right)^2 \tag{21}$$

where $m' = dm/d\rho$. The homogeneous contribution, $\mu\big(\rho(z)\big)$, can be calculated the same way as the free energy by using a mean-field EoS. Obviously, applications of Eq.(21) and the intrinsic free energy (the integrand of Eq.(11)) rely on an accurate expression for the density profile, $\rho(z)$, and numerical results from Eq.(18) or computer simulations require extra avenue for the derivatives.

Eq.(21) is an important result for exploring the interfacial region, but it is seldom addressed in the literature. The difficulty is that both the first and second derivatives of $\rho(z)$ are demanded reliably. Yong et al. [7] discussed it with some hypothetical scenarios without involvement of the density profile and any specific expression for $\mu(\rho(z))$. In this work, this equation will be employed for the Lennard-Jones fluid.

For applications of the density gradient theory, Eq.(11) and Eq.(21) etc., the density profile, $\rho(z)$, and it's derivatives, $d\rho/dz$ and $d^2\rho/dz^2$ are required and an analytical function can best serve the purposes. Several analytical expressions have been proposed [23] with the same dividing interface, Eq.(8). For a density profile model, the following boundary conditions should be imposed [1,3]:

$$\left.\begin{array}{c} z - z_0 = -\infty,\ \rho(z) = \rho_v \\ z - z_0 = \infty,\ \rho(z) = \rho_L \\ \left.\dfrac{d^n\rho(z)}{dz^n}\right|_{z\to\pm\infty} = 0 \end{array}\right\} \qquad (22)$$

A simple "classic" model that meets all the above conditions reads [1,3,8-12]:

$$\rho(z) = \frac{\rho_v + \rho_L}{2} + \frac{\rho_L - \rho_v}{2}\tanh\left(\frac{z - z_0}{D}\right) \qquad (23)$$

where the parameter $D$ is related to the thickness of the interfacial region for a planar surface [23]. Obviously, the algebraic-mean interface defined by Eq.(8) is adopted in Eq.(23). A fundamental issue is that the equal boundary "thicknesses" (decay length) at the vapor and the liquid sides given by Eq.(8) apparently contracts with the physical observations [6]. This issue has a big cost: all the position-dependent properties based on Eq.(23) are not acceptable, as shown in the next section.

In this work, based on a novel solution to the governing differential equation for the density profile, a different model is proposed (see Appendix A):

$$\rho(z) = A + \frac{B\tanh\left(\frac{z - z_0}{D_v}\right)}{1 + C\tanh\left(\frac{z - z_0}{D_L}\right)} \qquad (24)$$

where the coefficients $A$, $B$ and $C$ are dependent on the bulk fluid densities, $\rho_v$ and $\rho_L$. $z_0$, $D_v$ and $D_L$ are the parameters of the model to be determined from computer simulation data. It is straightforward to prove that if the conditions given by Eq.(8) and Eq.(22) are imposed to Eq.(24) the classic model, Eq.(23) is recovered, where $C = 0$ is a result of imposing the condition, Eq.(8).

An apparent feature of the new model is that two parameters, $D_v$ and $D_L$, are introduced to separate the vapor boundary from the liquid boundary. In addition, a new dividing interface is employed to incorporate with the separation. In solving a VLE problem, namely calculating the saturated densities and the equilibrium pressure at a given temperature, a mean-field EoS is usually employed. A vast majority of equations of state provide three solutions: one as the saturated vapor volume, another as the liquid volume, plus an "unphysical" solution. The last one has been considered "unphysical" since it corresponds to a state in the unstable interfacial region and has always been discarded in all the VLE calculations. In fact this "unphysical" solution, $\rho_M$, is well defined mathematically as a companion of the saturated volumes. Moreover, in the context of the density gradient theory, any state and related properties in the interfacial region can be meaningfully defined, such as those defined by Eq.(11) and Eq.(21). Some state, e.g. the average density, Eq.(8), may have no particular physical significance. Here we define the trajectory of the third solution, $\rho_M$, denoted as the M-line, as the mean-field dividing interface. The M-line or the Maxwell crossover can be generated from the Maxwell construction as detailed in Appendix B:

$$z - z_0 = 0,\ \ \rho(z) = \rho_M \qquad (25)$$

Combining Eq.(22) and Eq.(25) with Eq.(24) yields the new density profile:

$$\rho(z) = \rho_M + \frac{2(\rho_L - \rho_M)(\rho_M - \rho_v)\tanh\left(\frac{z - z_0}{D_v}\right)}{\rho_L - \rho_v + (2\rho_M - \rho_v - \rho_L)\tanh\left(\frac{z - z_0}{D_L}\right)} \qquad (26)$$

The parameters, $D_v$, $D_L$ and $z_0$ can be obtained by fitting the model with computer simulation data. Compared with the classic model, Eq.(23), the new model has one extra parameter and therefore is more flexible. Most importantly, the new model separates the vapor side of the interface from the liquid side and hence it is physically favorable. From the analytical expression, we can easily obtain the derivatives, $d^n\rho(z)/dz^n$. The first two derivatives to be used in this work are provided in Appendix A.

## The Lennard-Jones fluid

### A. The LJ EoS and related properties

The Lennard-Jones (LJ) fluid is one of the most important model fluids employed in various areas of condensed matter physics. The LJ fluid has been extensively studied with successful theories and computer simulations. For the vapor-liquid interface system, many publications have been devoted to the LJ fluids and their mixtures [8-11] or to the truncated and shifted LJ fluid [12,23]. For thermodynamic property descriptions including VLE calculations, a mean-field equation of state is usually employed. For a thorough and comprehensive review and comparisons of various EoS developed so far for the LJ fluid, the reader is refer to a recent publications (Stephan et al., 2020) [25,26]. The inter-molecular potential of the LJ fluid is given by the following:

$$u(r) = 4\epsilon\left[\left(\frac{\sigma}{r}\right)^{12} - \left(\frac{\sigma}{r}\right)^{6}\right] \qquad (27)$$

where $\epsilon$ and $\sigma$ are the energy and diameter parameters, respectively. First of all, we need density profile data for fitting the parameters of Eq.(26) and Eq.(23). Duque et al. (2004) [9] reported simulation data for the density profile at several temperatures. Very recently, Mejia et al. (2021) [8] published their simulation results along with Python code. In this study, the data in the temperature range, $6.5 \leq T^* \ll 1.1$, for the density profile [8,24] are employed and the parameters of Eq.(26) are correlated with temperature by polynomial functions:

$$D_L = 3.71168T^{*2} - 3.11467T^* + 1.77358 \qquad (28)$$

$$D_v = 6.79815T^{*2} - 8.85396T^* + 3.83015 \qquad (29)$$

$$z_0 = 44.82538T^{*3} - 90.6502T^{*2} + 69.70747T^* - 17.8015 \qquad (30)$$

For the classic model, Eq.(23), it's parameters are also obtained from the same data and correlated with the following equations:

$$D = 9.14288T^{*3} - 19.1378T^{*2} + 15.0215T^* - 3.14391 \qquad (31)$$

$$z_0 = 43.7861T^{*3} - 87.9168T^{*2} + 66.8356T^* - 15.5913 \qquad (32)$$

For calculations of various thermodynamic properties here we adopt the mean-field LJ EoS proposed recently by Stephan et al. [25,26] due to it's high accuracy and relative simplicity. In all the calculations, reduced quantities are used and indicated with the superscript "*". For example, $\rho^* = N\sigma^3/V$, $T^* = T/k_B\epsilon$, $P^* = P\,\sigma^3/k_B\epsilon$, $a^* = F/\epsilon$, $\Gamma^* = \Gamma\sigma^2$, $\gamma^* = \gamma\sigma^2/\epsilon$ etc. In the mean-field EoS, the residual (relative to the ideal gas) Helmholtz free energy is composed of a reference term and a perturbation contribution [25]:

$$a^* = T^*(ln\rho^* + \tilde{a}^{res}) = T^*(ln\rho^* + \tilde{a}^{ref} + \tilde{a}^{pert}) \qquad (33)$$

The term $ln\rho^*$ comes from the ideal gas and a temperature-dependent term is ignored without impacting the final results. The expressions for the reference term (the hard sphere) $\tilde{a}^{ref}$, the perturbation term $\tilde{a}^{pert}$ and related quantities can be found in Refs [25,26] and some details are given in Appendix B. The reduced pressure is given by:

$$P^* = \rho^* T^*\left(1 + \frac{\partial\tilde{a}^{res}}{\partial\rho^*}\right) \qquad (34)$$

and the chemical potential (the Gibbs free energy) is given by:

$$\mu^* = \frac{\partial\rho^*\tilde{a}}{\partial\rho^*} = T^*\left(ln\rho^* + \tilde{a}^{res} + 1 + \frac{\partial\tilde{a}^{res}}{\partial\rho^*}\right) \qquad (35)$$

With the above EoS, we have an analytical expression for the grand potential:

$$\Delta\Omega = \frac{1}{2}[p_N(z) - p_T(z)] = \left[a_0\big(\rho(z)\big) - \mu^{*s}\right]\rho(z) + P^{*s} \qquad (36)$$

where $a_0\big(\rho(z)\big)$ is the homogeneous contribution given by the mean-field EoS. Finally, the intrinsic Gibbs free energy can be calculated with the following:

$$\mathbf{G}^*[\rho(z)] \approx T^* ln\rho^* + T^*\tilde{a}^{res} + \frac{P^*}{\rho^*} - m\frac{d^2\rho}{dz^2} \qquad (37)$$

where the last term with $m'$ in Eq.(21) turned out to be negligible for the LJ fluid. The intrinsic local free energy is calculated from Eq.(11):

$$\boldsymbol{f}[\rho(z)] = \frac{f_0(\rho(z))}{\rho(z)} + \frac{1}{2}\frac{m}{\rho(z)}\left(\frac{d\rho}{dz}\right)^2 = T^*(ln\rho^* + \tilde{a}^{ref} + \tilde{a}^{pert}) + f_2(\rho(z)) \quad (38)$$

where $f_2(\rho(z)) = \frac{1}{2}\frac{m}{\rho(z)}\left(\frac{d\rho}{dz}\right)^2$ is the heterogeneous contribution and the density $\rho(z)$ is introduced to convert the free energy density to the reduced value. By recalling that all the above equations, Eq.(36)-Eq.(38), are exact up to the 3rd order while omitting the 4th order and higher terms, the results presented below are accurate enough for our analysis.

In addition to the intrinsic Helmholtz free energy and the Gibbs free energy, here we are also interested in heat capacity. The basic assumption is that classic thermodynamic definitions (formalisms) also apply to the derivative properties (or response functions) as the heterogeneous contribution is considered separately. Hence, for the vapor-liquid interface system, the intrinsic local heat capacity at constant pressure can be calculated from an EoS by [25]:

$$\boldsymbol{C_P}(\rho(z)) = \boldsymbol{C_v}(\rho(z)) + \frac{\left(1 + \tilde{a}_{01}^{int} - \tilde{a}_{11}^{int}\right)^2}{1 + 2\tilde{a}_{01}^{int} + \tilde{a}_{02}^{int}} \quad (39)$$

where the intrinsic heat capacity at constant volume is given by:

$$\boldsymbol{C_v}(\rho(z)) = -\left(\tilde{a}_{20}^{id} + \tilde{a}_{20}^{int}\right) \quad (40)$$

In Eq.(39) and Eq.(40), the derivatives are defined by the following:

$$\tilde{a}_{nm}^{int} = \tau^n \rho^{*m} \frac{\partial^{n+m}\left(\tilde{a}^{res} + \tilde{f}_2\right)}{\partial\tau^n \partial\rho^{*m}} \quad (41)$$

where $\tilde{a} = a/T^*, \tau = 1/T^*,\ \tilde{a}_{20}^{id} = 1.5$ and

$$\tilde{f}_2(\rho(z)) = \frac{1}{2T^*}\frac{m}{\rho(z)}\left(\frac{d\rho}{dz}\right)^2 \quad (42)$$

The homogeneous contribution to the residual free energy is defined as $\tilde{a}^{res} = a_0(\rho(z)) - \tilde{a}^{id}$, and the contribution of the gradient term in Eq.(38), should be included for evaluating the heat capacity, $\tilde{a}^{int} = \tilde{a}^{res} + \tilde{f}_2$. More details are provided in Appendix B.

At a given temperature, the saturated densities of the liquid and vapor phases and the saturated pressure can be calculated by the Maxwell construction [27], or equivalently by the pressure and chemical potential equilibrium conditions (Appendix B). In particular, the third solution ($v_M$) is obtained from the pressure equilibrium condition:

$$P^*(v_M) = P^*(v_L) = P^*(v_G) = P^{*S} \quad (43)$$

The state at $v_M$ is unstable since $\mu^*(v_M) \neq \mu^{*S}$. However, as shown above, Eq.(37) and Eq.(38), in the framework of the density gradient theory any state in the interfacial region is physically defined.

After the saturated densities and pressure are obtained over the entire coexistence region, they can be correlated as functions of temperature with some empirical equations. Such correlations can be found in Ref.[25,26] and adopted in this work to determine the M-line (Appendix B). The M-line turns out to be a simple linear function in the temperature range considered:

$$\rho_M = 0.3073T^* - 0.09778 \quad (44)$$

For applying Eq.(37), Eq.(38) etc., the influence parameter, $m$, is required. It can be estimated from surface tension via Eq.(16). As mentioned, this parameter is a weak function of temperature or density. Duque et al. [9] reported a constant value for the LJ fluid: $m = 4.4$. Since the derivative, $m' = dm/d\rho$, is required in Eq.(21), we adopt the values reported by Galliero et al. (2009) [11], which are between 5.2 and 5.5. Now we need to calculate the density derivative, $dm/d\rho$. In the coexistence region, there are two saturated densities at a given temperature for the liquid and vapor phases, respectively. Choosing either of them is not suitable since the derivatives $dm/d\rho$ will have opposite signs. On the other hand, we know that the influence parameter can also be expressed as a function of temperature, $m(T^*)$ and that $dm/dT^* < 0$, $dm/d\rho < 0$ [1,3,6,11]. By taking Eq.(44) into account, we can treat $m$ as a function of $\rho_M$ and we fitted the influence parameter data [11] with the following equation:

$$m = -2\rho_M + 5.80 \quad (45)$$

which gives $dm/d\rho = dm/d\rho_M = -2$.

Now we turn to the supercritical region. Numerous studies on various supercritical fluids can be found in the literature [14-22]. One of the most important findings is that the Widom line [13] divides a supercritical area into liquid-like and gas-like regions. Some other characteristic lines are also found. For a detailed discussion of the relationships between different characteristic lines, the reader is referred to a recent publication (Banuit et al., 2020) [17]. Here we are only interested in the region where the Widom line functions as the dividing curve. The Widom line is defined as the loci of local maximum of the isobaric heat capacity, namely [19]:

$$\left(\partial C_p/\partial T\right)_P = 0 \tag{46}$$

By using the LJ EoS [25], calculations are carried out at various conditions with Eq.(39) and (Eq.(46). For the interfacial region, investigations are focused in the temperature range, $0.65 \leq T^* \leq 1.0$.

## B. Calculation results and discussions

All calculation results are presented by Figure 1 – Figure 11, where the position variable is denoted as $x = z - z_0$ for brevity. Figure 1a depicts definitions of the classic (the algebraic mean) dividing interface, Eq.(8) and the mean-field interface, Eq.(25). In this work, the thickness of the interface will not be addressed and the interface area (between the dashed blue lines) is illustrative only. The mean-field dividing interface correctly reflects the fact that the boundary layer at the vapor side is thinner than that at the liquid side. In comparison, the algebraic mean interface equally (wrongly) divides the interface area.

Figure 1b presents the 1st and 2nd derivatives obtained from the mean-field density profile, Eq.(26) and the classic density profile, Eq.(23) at $T^* = 0.65$. The same behavior is found at other temperatures. This is a very important comparison. The positional dependences obtained from the two models are different. Apparently, the positional differences are caused by the different definitions of the dividing interface. Because Eq.(26) (with Eq.(28)-Eq.(30)) is more accurate than Eq.(23) (with Eq.(31)-Eq.(32)), as shown by Figure 2, the derivatives obtained from the classic model are inaccurate. From Eq.(37) and Eq.(38) etc., we know that predictions of the intrinsic properties from the classic model, Eq.(23), are not reliable. This also explains the failure of the classic model in prediction of the positional dependence of the pressure difference, $p_N(z) - p_T(z)$ (see Ref.[12]).

Figure 2 illustrates detailed correlation results for the density profiles over the entire temperature range, $0.65 \leq T^* \leq 1.1$, by the classic model, Eq.(23) (dashed lines), and by the new model, Eq.(26) (solid red lines). The parameters are given by Eq.(28)-Eq.(30) for the new model, and by Eq.(31) and Eq.(32) for the classic model. The observations are: (1) the new model, Eq.(26), works excellently over the entire range from vapor phase to liquid phase; (2) the classic model, Eq.(23), is less accurate compared to Eq.(26), in particularly on the deep liquid side (Figure 2c).

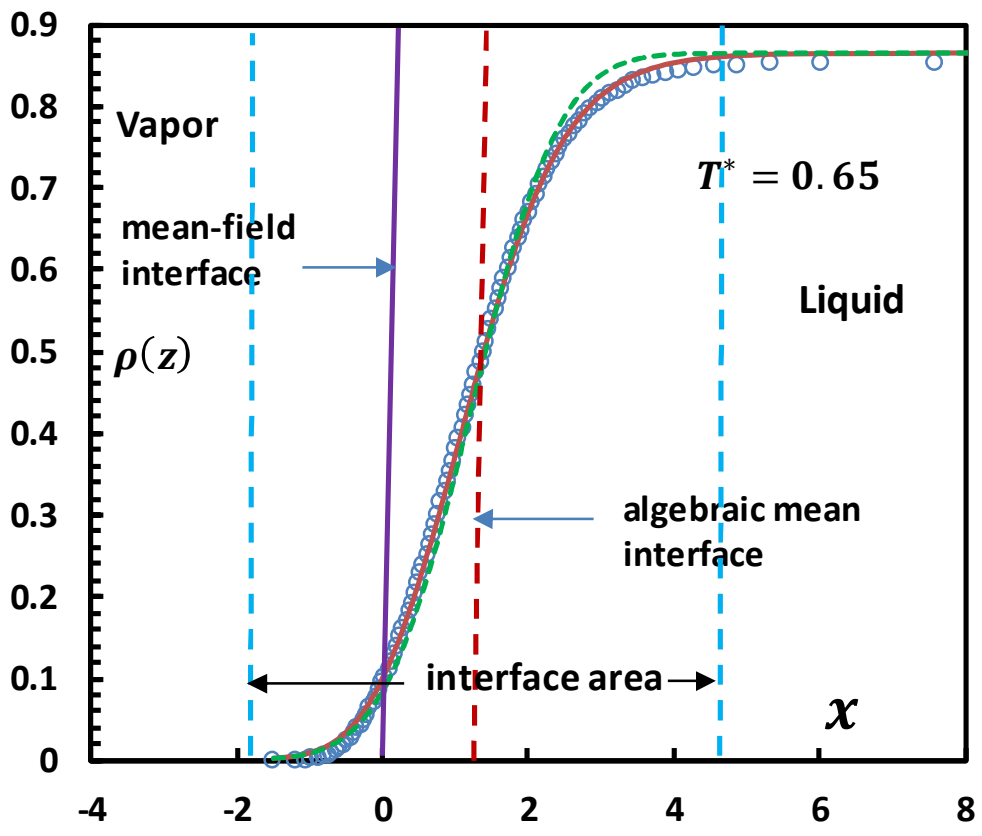

**Figure 1a**. Density profile at $T^* = 0.65$, where $x = z - z_0$. The interfacial region is defined illustratively between two dashed blue lines. The red solid curve is from Eq.(26), dashed green line from Eq.(23) and circles are from the most recent simulation data [8,24].

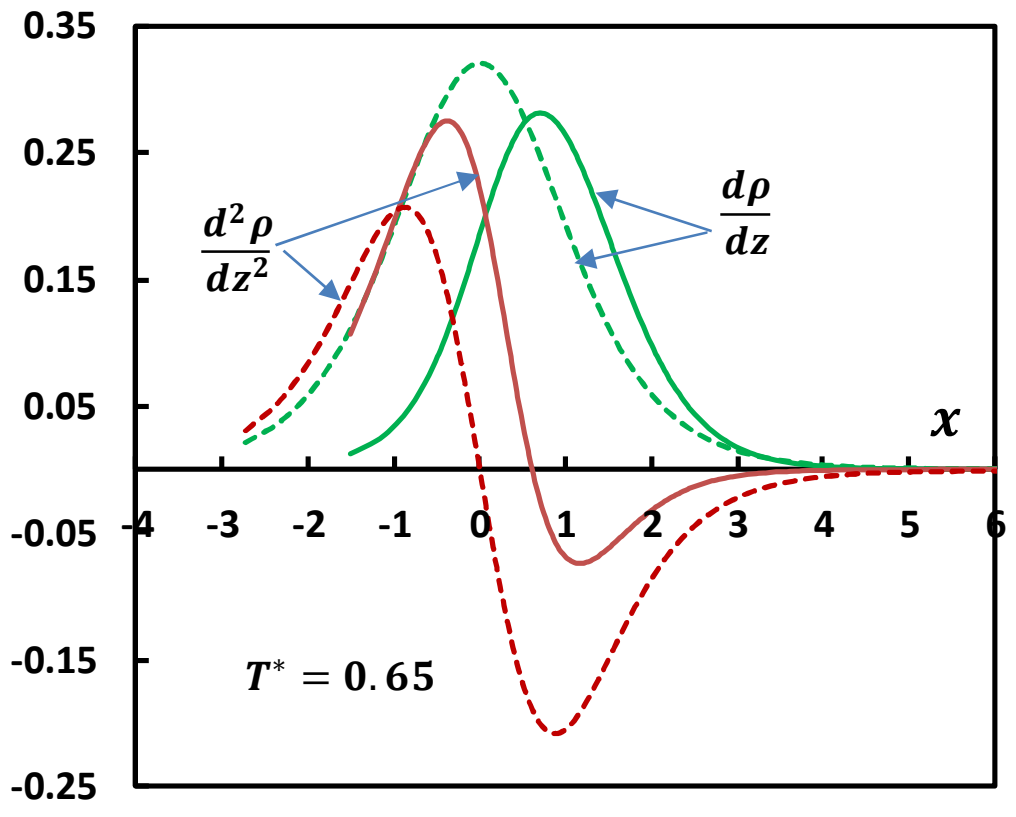

**Figure 1b**. Derivatives from two density profile models: solid lines are from Eq.(26) and dashed lines from Eq.(23).

**Figure 1**. Density profiles and the derivatives at $T^* = 0.65$

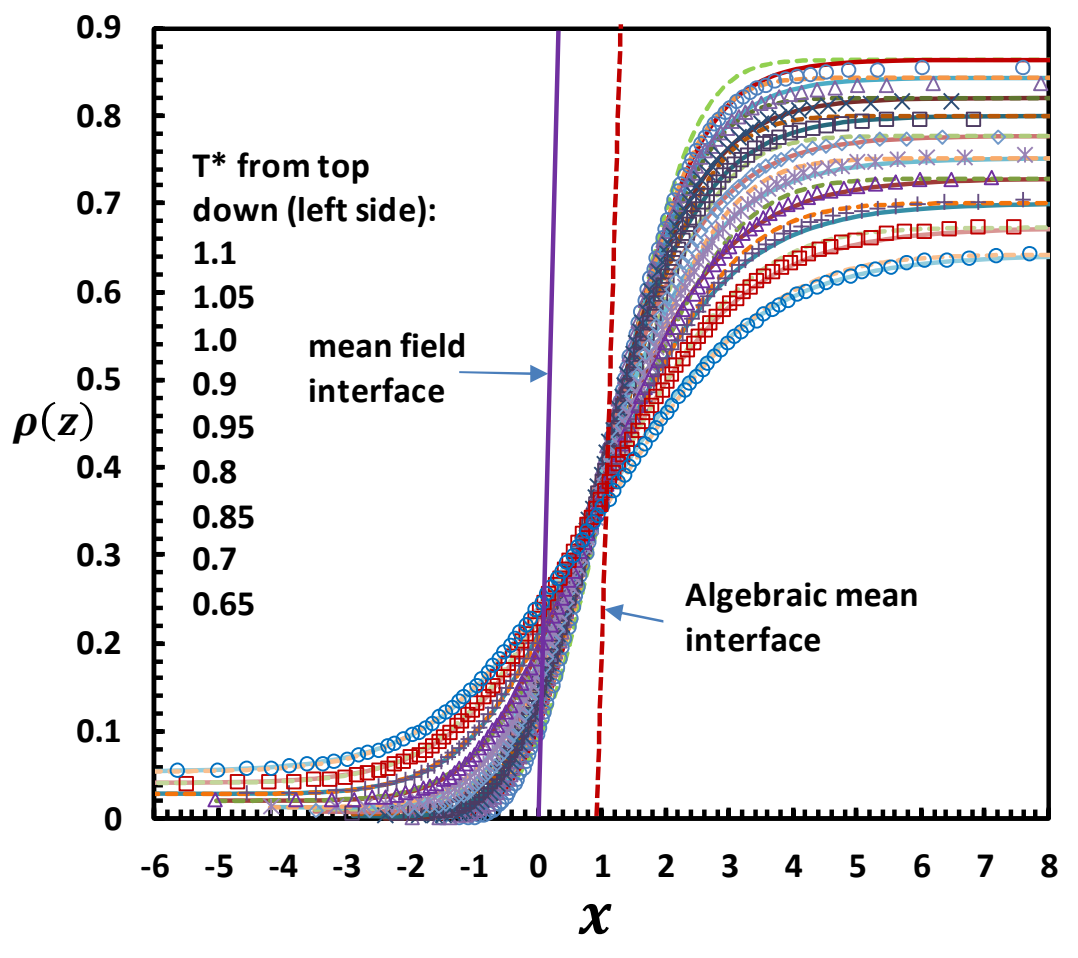

**Figure 2a**. Plots of density profiles in the entire range.

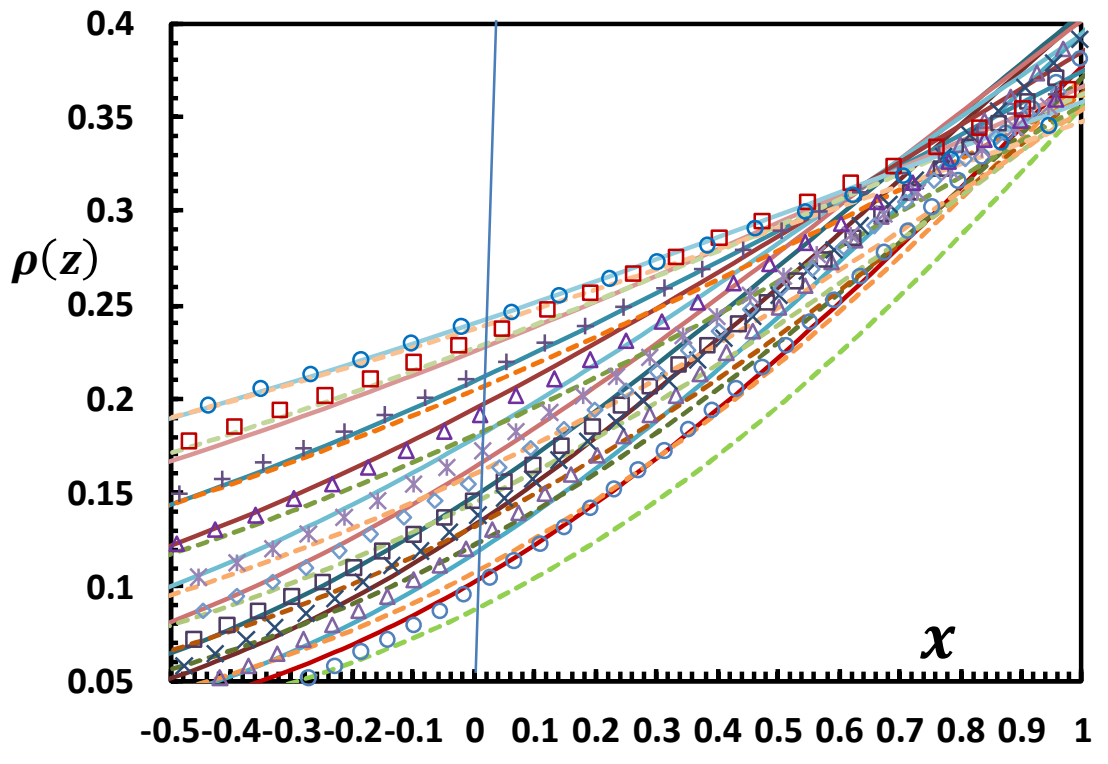

**Figure 2b**. Plots of density profile correlations: the magnified portion around the dividing interface.

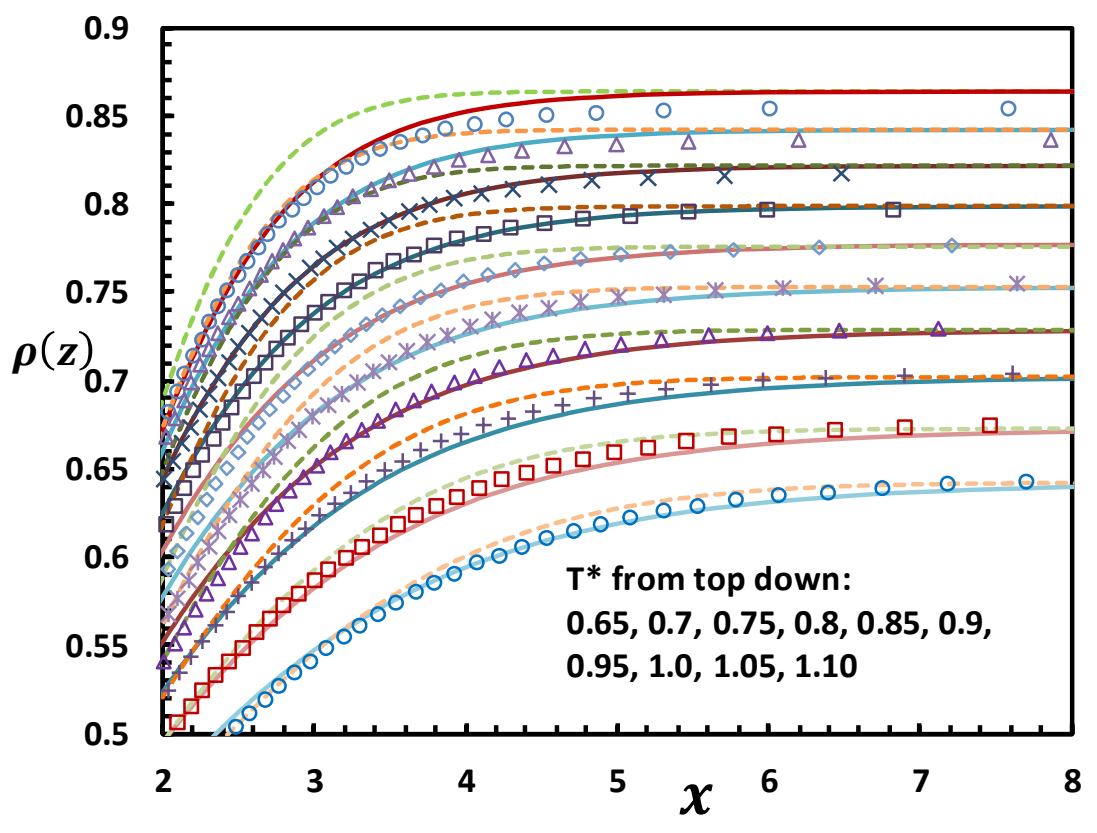

**Figure 2c**. Plots of density profile correlations: enlarged portion on the liquid side.

**Figure 2**. Plots of density profile correlations. Solid lines: Eq.(26); dashed lines: Eq.(23); points: Ref. [8,24].

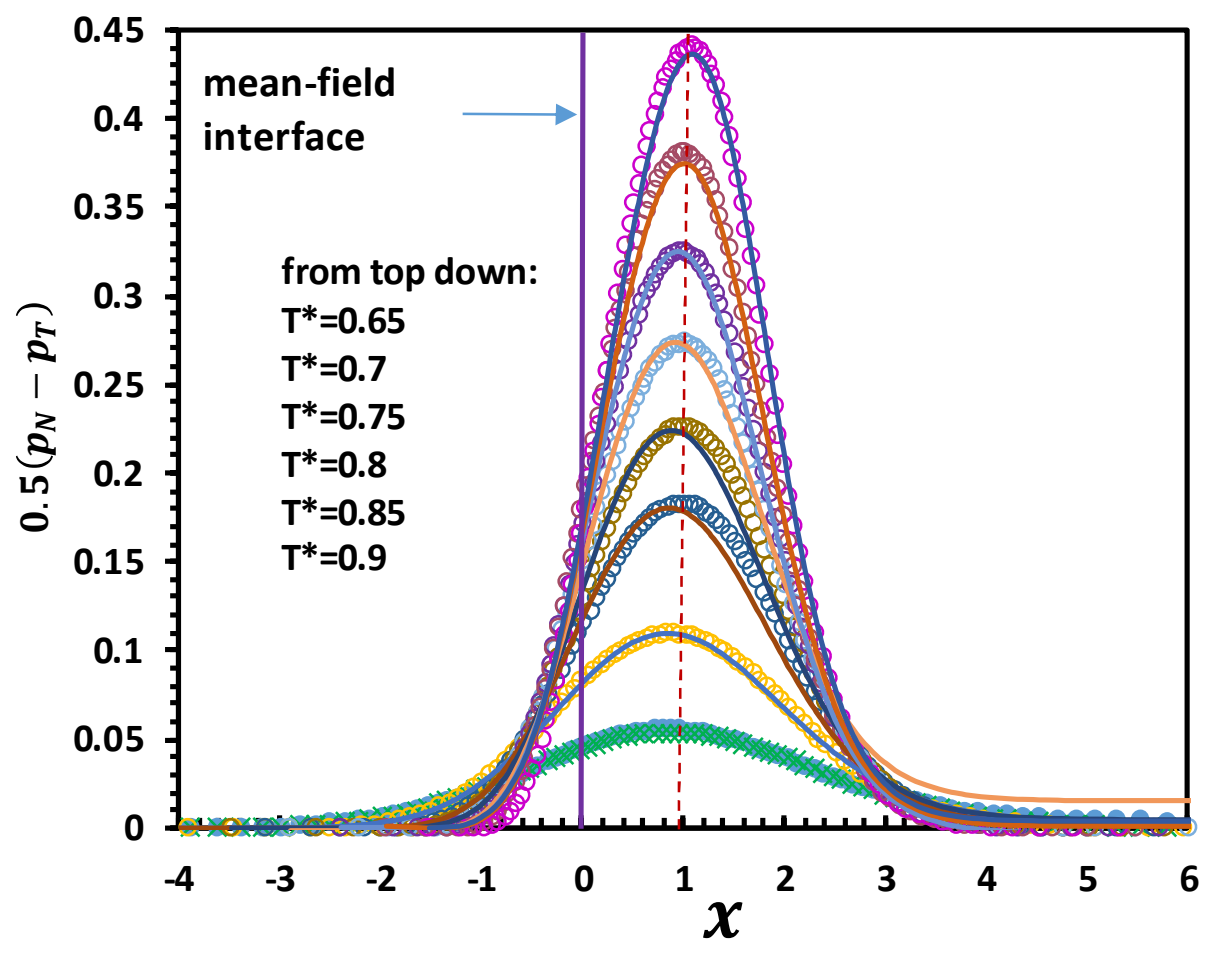

**Figure 3**. Plots of predictions of the pressure difference, $[p_N(z) - p_T(z)]/2$ by Eq.(36) and Eq.(26) (lines), compared with the simulation data [24].

Figure 3 presents the prediction results for the pressure difference, $p_N(z) - p_T(z)$, by Eq.(26) combined with Eq.(36). The simulation values and predictions of Eq.(26) show that $p_N(z) - p_T(z)$ peaks at $x \approx 1$. The results illustrated by Figure 2 and Figure 3 demonstrate that the mean-field theory works excellently with the density profile, Eq.(26). By the way, as shown by Eq.(4), the maximum of the pressure difference implies that the change rate (w.r.t position) of the local surface tension peaks at the same position, $x \approx 1$. The high correlation and prediction accuracies shown by Figure 2 and Figure 3 are of crucial importance for applying Eq.(37) and Eq.(38) in succeeding calculations.

Figure 4a depicts the Gibbs free energies across the interface area at $T^* = 0.85$. It shows that at $x = z - z_0 \approx 1$, there exhibits a maximum with the intrinsic Gibbs free energy, Eq.(37) with Eq.(26), while the classic model, Eq.(23) (with Eq.(37)), fails to predict the behavior. This observation echoes the results shown by Figure 1b. Another important observation is that the Maxwell construction or the equal-area rule holds for the homogeneous contribution where the density $\rho(z)$ is used in the LJ EoS. In other words, the mean-field formalisms hold for various thermodynamic properties with the density $\rho(z)$, while heterogeneous contribution is covered separately by Eq.(42). It is found that in Eq.(37) the term $\frac{1}{2}m'\left(\frac{d\rho}{dz}\right)^2$ can be neglected without impacting the final results. Therefore the impact of density-dependence of the influence parameter is indeed minor.

Figure 4b illustrates the intrinsic Gibbs free energy at different temperatures and it is found that all curves peak at $x \approx 1$. By comparing Figure 4b with Figure 3, we see that the maxima of the pressure difference, $p_N(z) - p_T(z)$, correspond to the maxima of the intrinsic Gibbs free energy (at the liquid side). For a stable fluid Eq.(3) shows that the surface tension is directly related to the Gibbs free energy macroscopically. In comparison, Eq.(4), Figure 3 and Figure 4b demonstrate that the intrinsic Gibbs free energy is related to the change rate of the local surface tension.

At the vapor side, the magnitude of pressure is much less than that at the liquid side and the pressure difference, $p_N(z) - p_T(z)$, is hard to detect, therefore a similar correspondence does not apply. Figure 4 also shows that for the LJ fluid, the intrinsic Gibb free energy exhibits a minimum at the interface, $x = 0$.

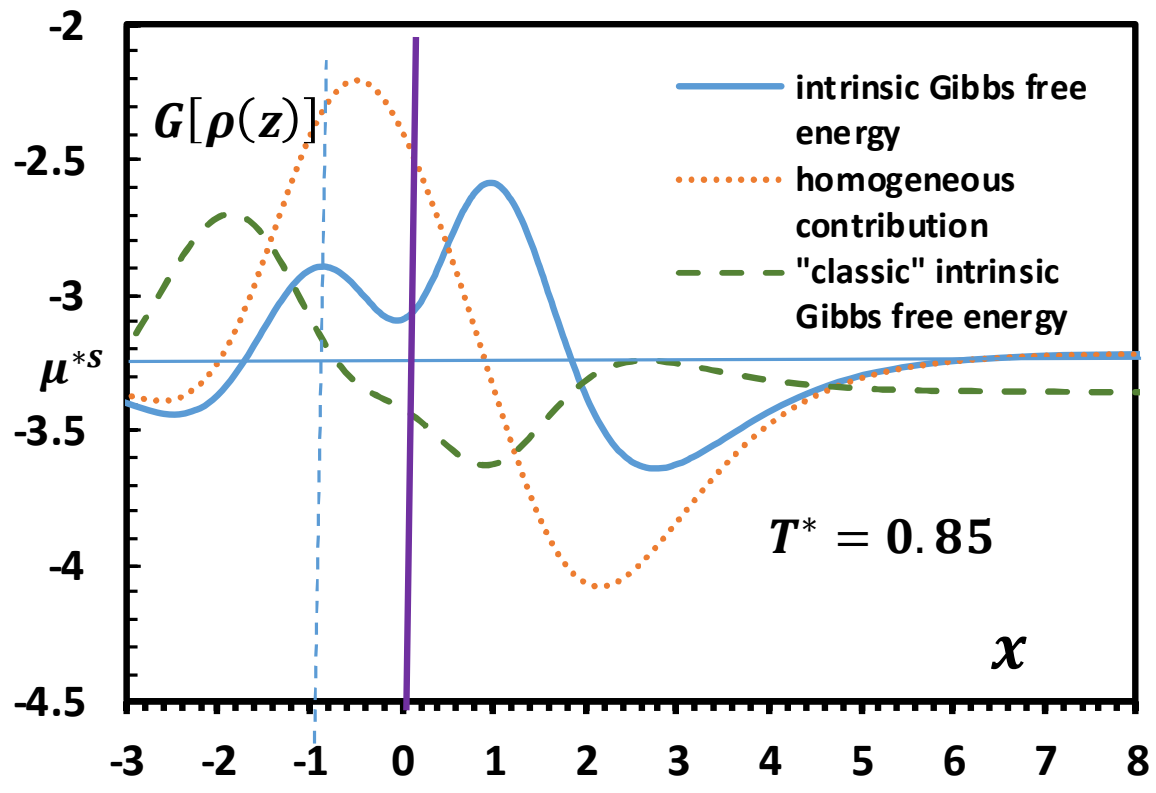

**Figure 4a**. A comparison of various Gibbs free energies at $T^* = 0.85$. The solid blue line is from Eq.(37) with the density profile from Eq.(26), and the dashed green line from Eq.(23). The dotted line shows the contribution from the homogeneous mean-field EoS, $\mu^*(\rho(z))$. The horizontal solid line represents the equilibrium value, $\mu^{*s}$.

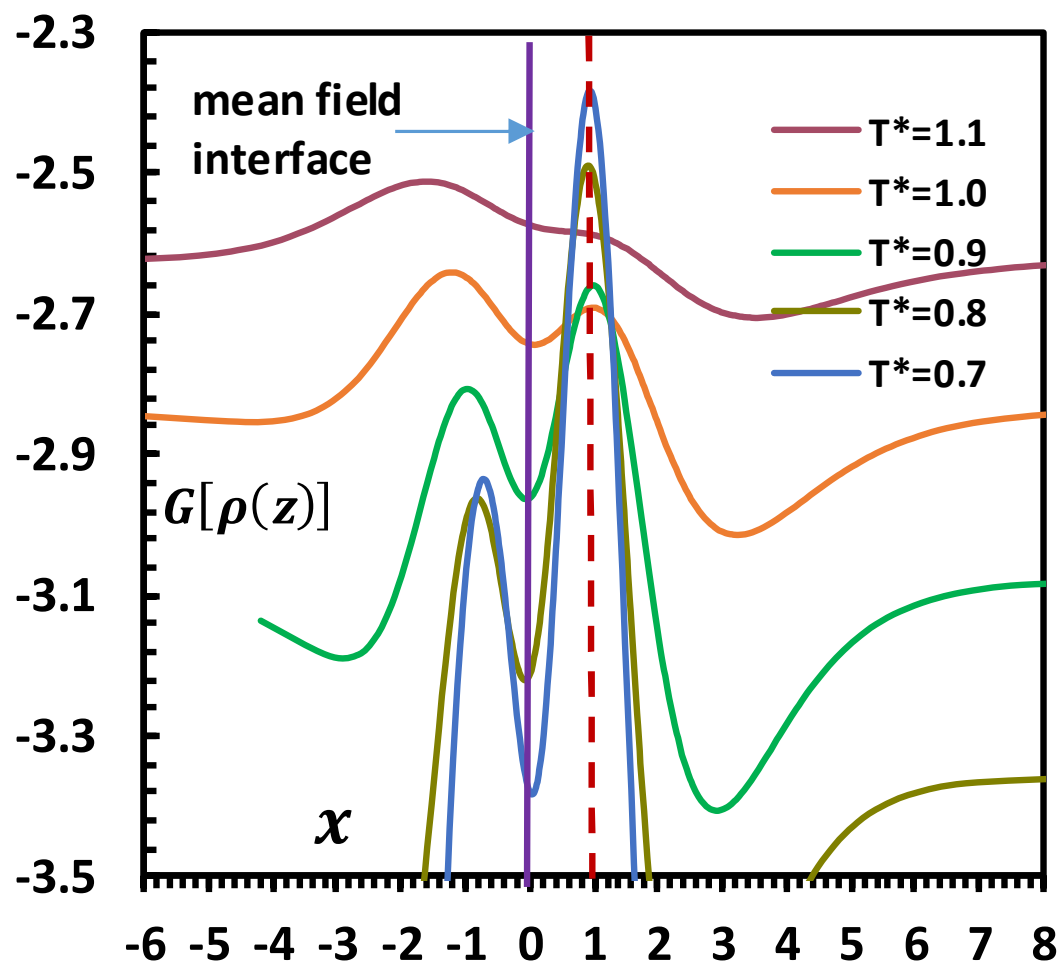

**Figure 4b**. The intrinsic Gibbs free energy calculated by Eq.(37) at different temperatures.

**Figure 4.** Plots of the Gibbs free energy.

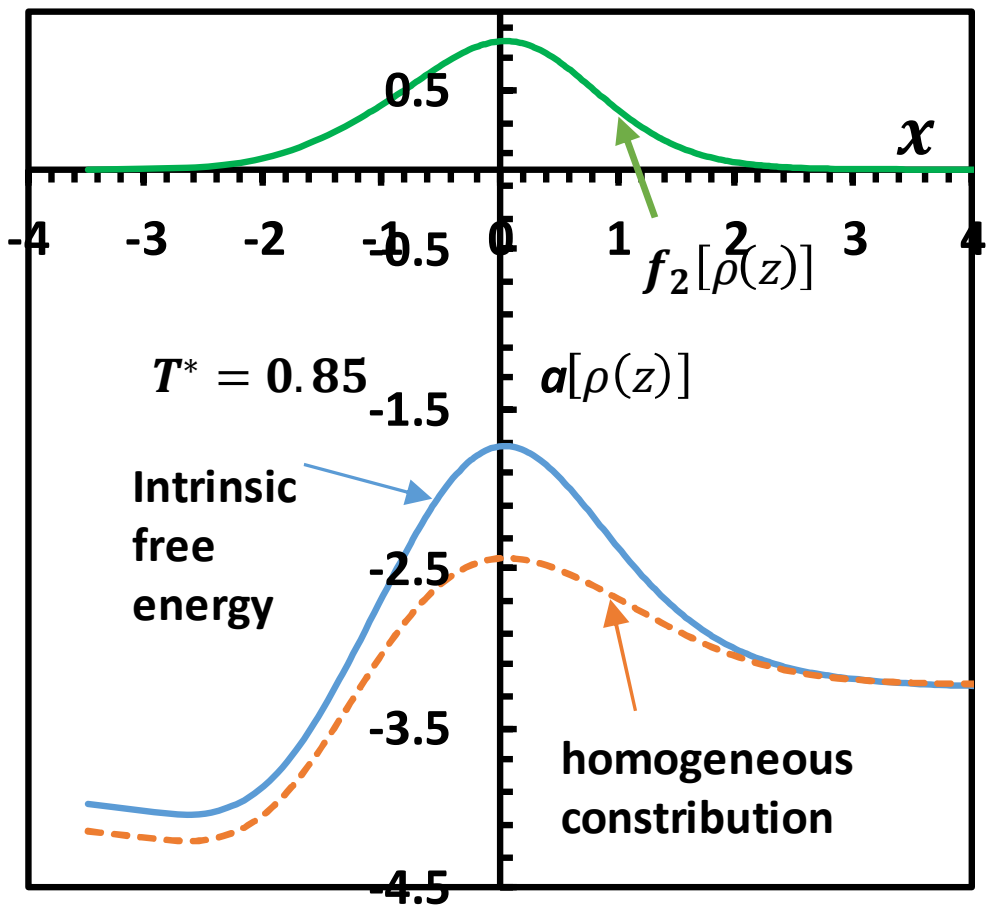

**Figure 5**. Plots of the Helmholtz free energies at T*=0.85. The intrinsic free energy (solid blue line) is calculated by Eq.(38). The homogeneous contribution is from Eq.(33) with the density $\rho(z)$ by Eq.(26). The green line is the heterogeneous contribution.

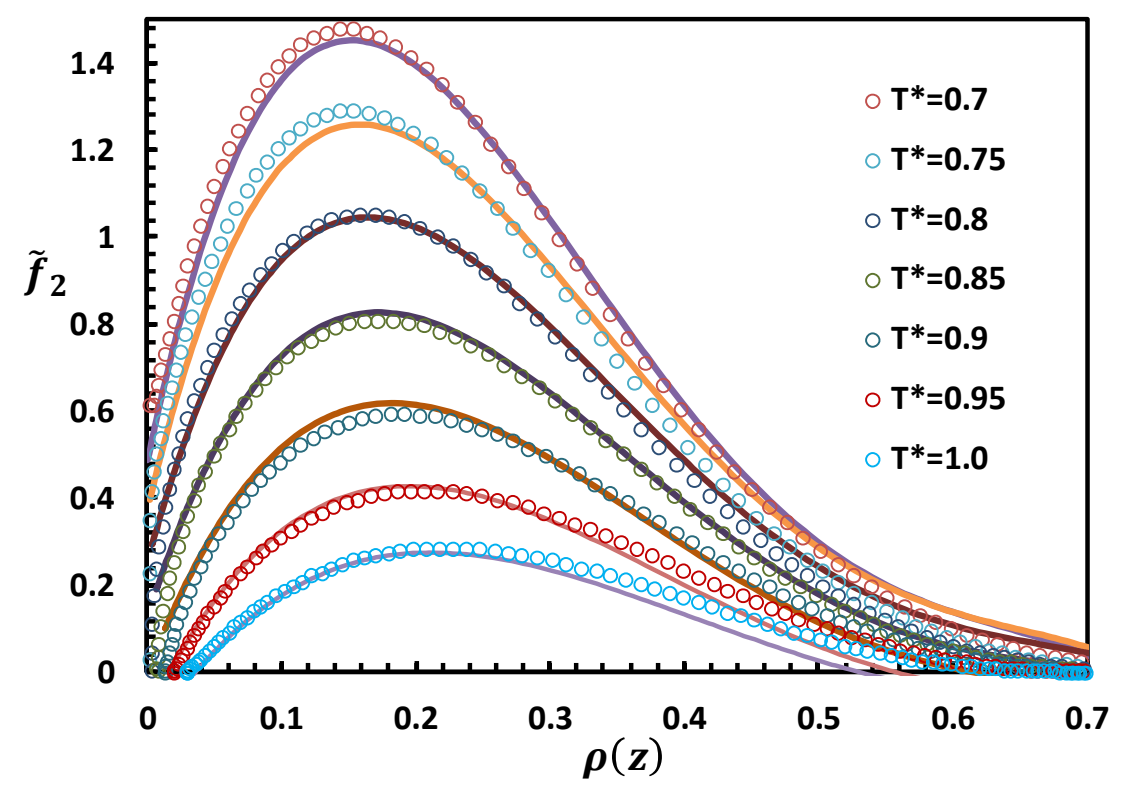

**Figure 6**. Heterogeneous contributions of the reduced fee energy, $\tilde{f}_2$ defined by Eq.(42). The points are calculated from simulation data of the density profile [24], and lines are correlations from the generic correlation, Eq.(B14) and Eq.(B15) (Appendix B).

Figure 5 depicts the free energy at $T^* = 0.85$. The intrinsic free energy peaks at the dividing interface, namely a state on the interface is the most unstable one. The local maxima of the isobaric heat capacity is in accordance with the behavior of the free energy, as discussed below. This feature is a result of choosing the mean-field crossover as the dividing interface. As shown by Figure B6, when the classic density profile, Eq.(23), is used, the local maximum location of the free energy and heat capacity does not coincide with the dividing interface (Appendix B). Figure 1b explains the reason. The later demonstrates from a different perspective that using the M-line as the dividing interface is physically favorable.

Figure 6 presents the heterogeneous contributions of the reduced fee energy, $\tilde{f}_2$ defined by Eq.(42) over a temperature range. At high temperature the heterogeneous contribution plays a weaker role than it does in low temperature. This causes the heat capacity behaving differently below and above $T^* \approx 0.85$ ($T_r \approx 0.65$): as $T^* > 0.85$, $C_P < 0$ and as $T^* < 0.85$, $C_P > 0$ (see Figure 7).

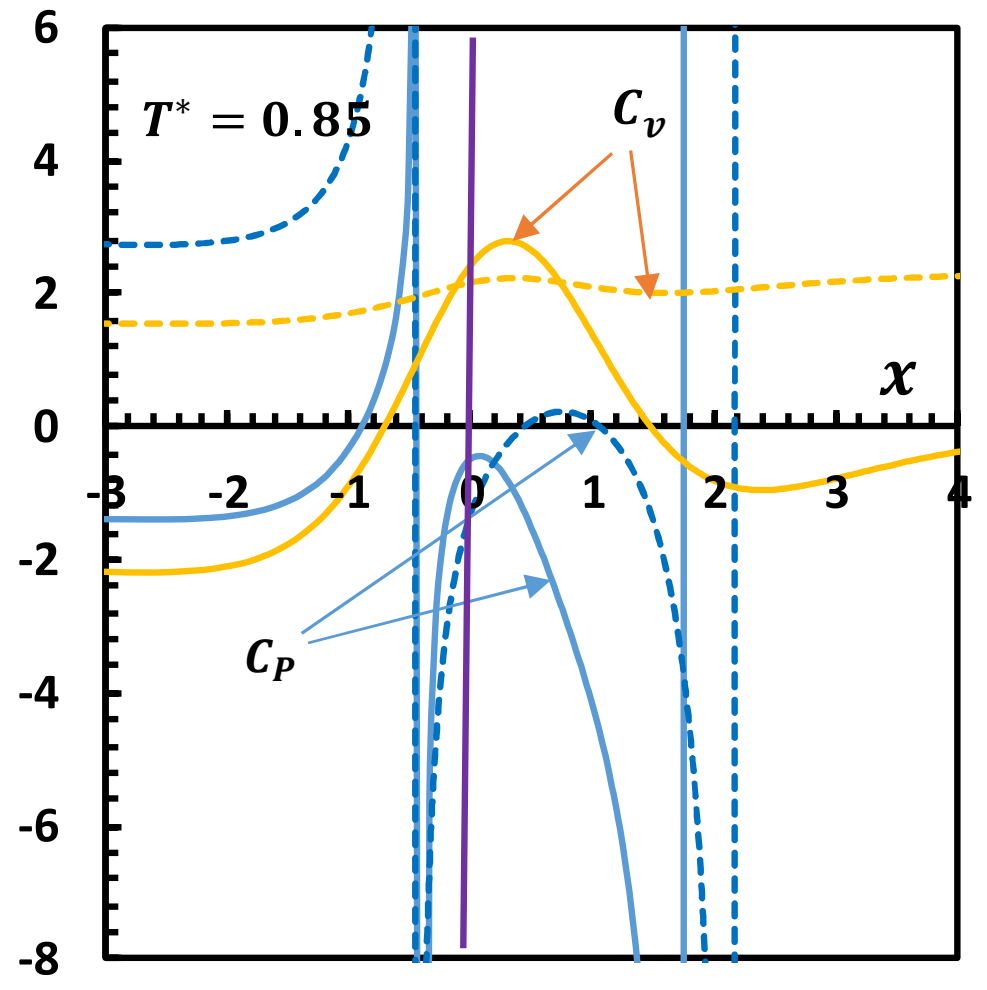

**Figure 7a** Plot of heat capacities at T*=0.85. The solid lines represent the intrinsic heat capacities and dashed line depict the homogeneous contributions.

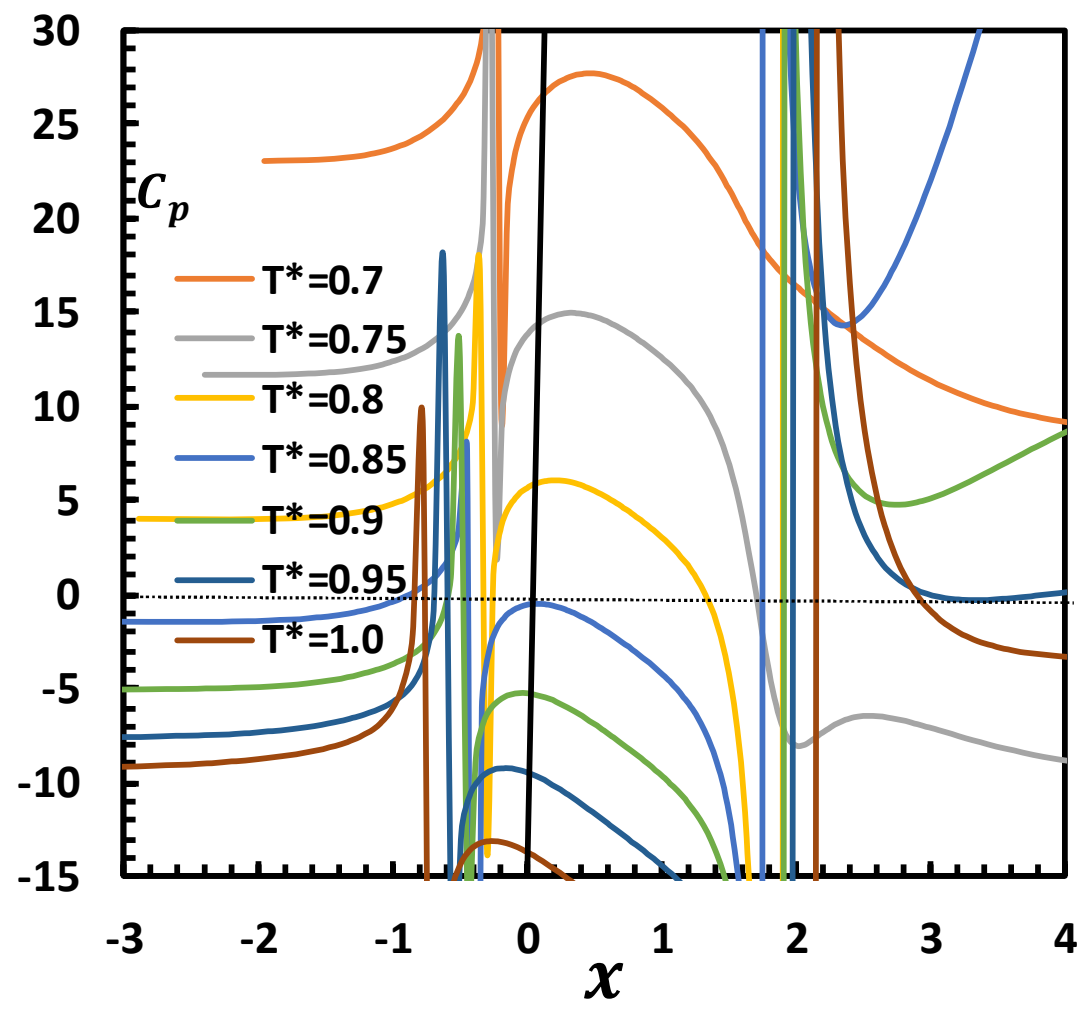

**Figure 7b**. Intrinsic isobaric heat capacity at multiple temperatures.

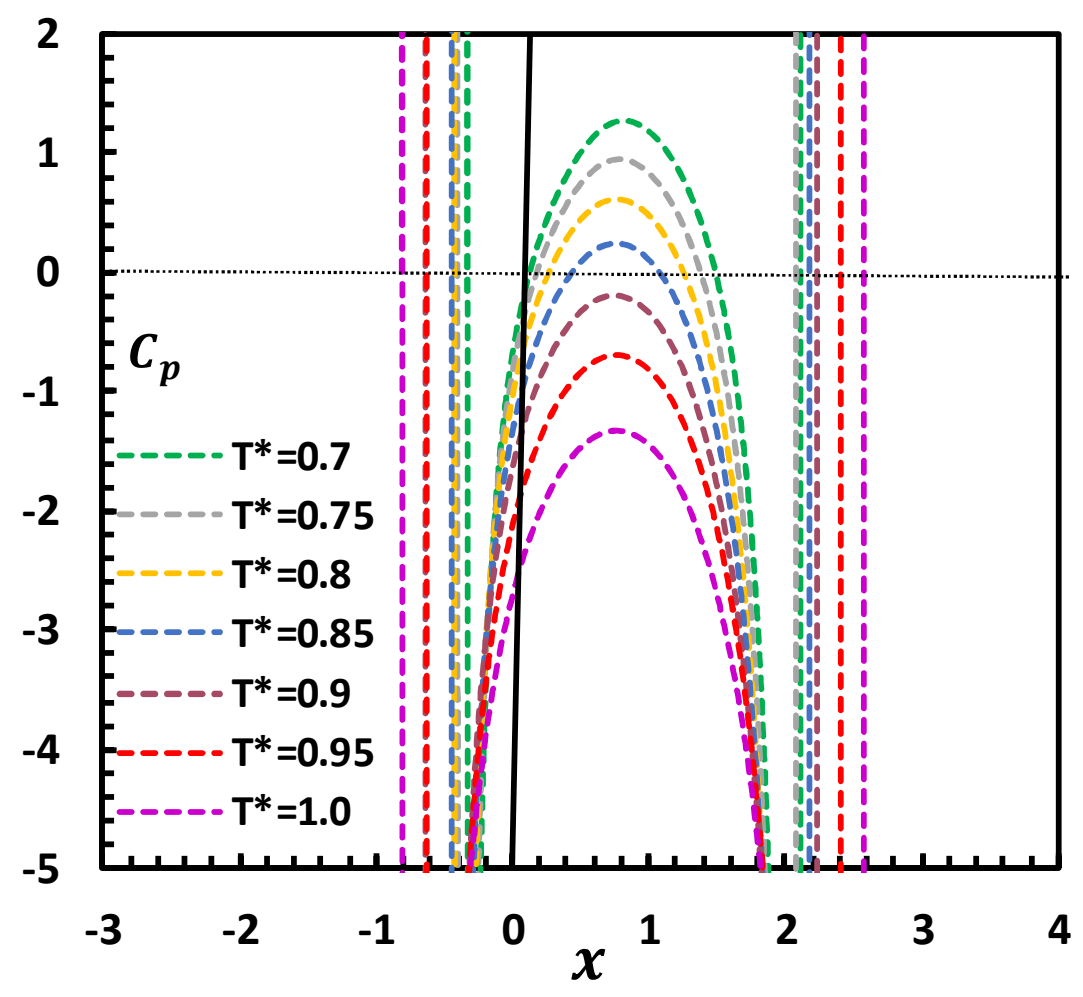

**Figure 7c.** Homogeneous contributions of the isobaric heat capacity at several temperatures.

**Figure 7**. Plots of local heat capacity for the interfacial region.

Figure 7a depicts the isobaric and isochoric heat capacities calculated by Eq.(39) and Eq.(40), respectively, at T*=0.85. At this temperature the maximum of $C_P$ approximates zero. It is clearly shown that $\tilde{f}_2$ determines the position of the maximum. In the stable fluid region, $C_P > C_v > 0$, while in the interfacial region, $C_P < C_v$. This implies that in the interface area temperature has a profound impact on the internal energy.

Figure 7b illustrates the intrinsic isobaric heat capacity at different temperatures. Figure 7c depicts the homogeneous contributions to the heat capacity at those temperatures, which shows that the sign becomes positive as $T^* < 0.85$. The most important observation from Figure 7b is that the intrinsic heat capacity $C_P$ peaks (local maximum) at the mean-field interface. Some small deviations of the maxima from the interface are mainly caused by the discrepancies between the saturated densities from Ref.[8,24] (for $\rho(z)$ and $\tilde{f}_2$) and from Ref.(25] (the LJ EoS for $C_v$ and $C_P$). Some details are provided in Appendix B. In addition, in high ($T^* \geq 1.0$) and low ($T^* \leq 0.7$) temperature regions the inaccuracies of the generic correlation, Eq.(B14), also contribute to the deviations. Another interesting observation is that at low temperatures ($T^* < 0.85$), $C_v > 0$ and $C_P > 0$. This finding is unexpected since previous theories or observations argue that in heterogeneous or nanoscale systems the heat capacity is negative [29-31]. As temperature rises ($T^* \geq 0.85$), negative heat capacity is indeed observed. The calculation details and more results for the heat capacity can be found in Appendix B.

In summary, with the mean-field dividing interface the density gradient theory, incorporated with the mean-field theory, predicts the following interfacial behaviors: (1) maximum intrinsic Helmholtz free energy at the interface; (2) maximum Gibbs free energy in the liquid side, in accordance with the maximum pressure difference, $p_N(z) - p_T(z)$; (3) local maximum of the intrinsic isobaric heat capacity at the interface with positive values in low temperature region, $T_r^* < 0.65$, and negative values in high temperature region, $0.65 < T_r^* < 1.0$. These features are not found by using the classic density profile model, Eq.(23) (Figure B6).

Now we move to the supercritical region by using Eq.(46) to generate the Widom line [13,14,19]. Figure 8 depicts the isobaric heat capacity in the supercritical region for the LJ fluid calculated with Eq.(39)-Eq.(42). The Widom line can be obtained by following a constant pressure process: (1) at a given pressure in the supercritical region, calculating the heat capacities for a temperature range in which a maximum of $C_p$ is observed; (2) from the $C_p \sim T$ plot, finding the maximum point, $T_m^*$, and solving the LJ EoS [25] to obtain the density at $T_m^*$; (3) repeating the calculations at another pressure.

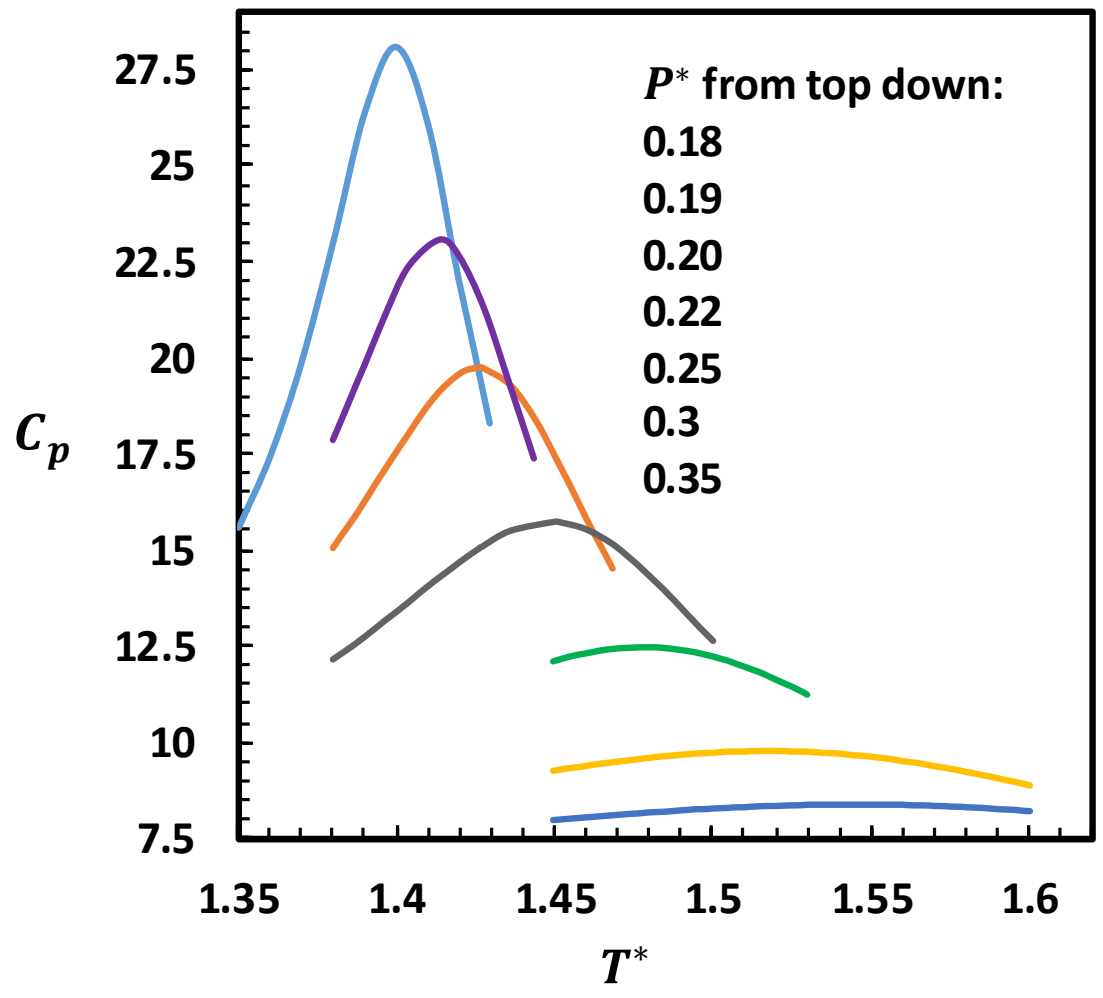

**Figure 8**. Determination of the locations of the maxima of the isobaric heat capacity from Eq.(46) in the supercritical region.

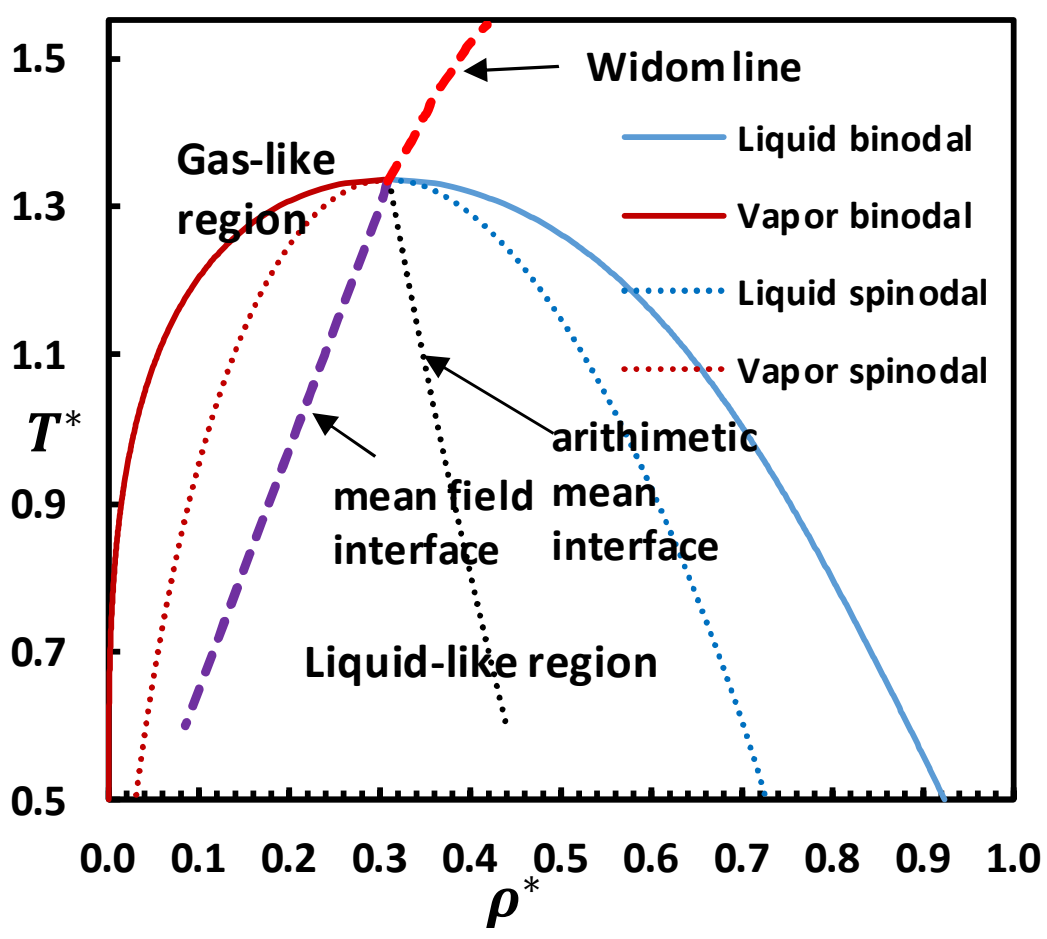

**Figure 9**. Phase diagram in the temperature-density $(T^*, \rho^*)$ plane. The spinodal and binodal curves are produced from the data reported in Refs.[25,26].

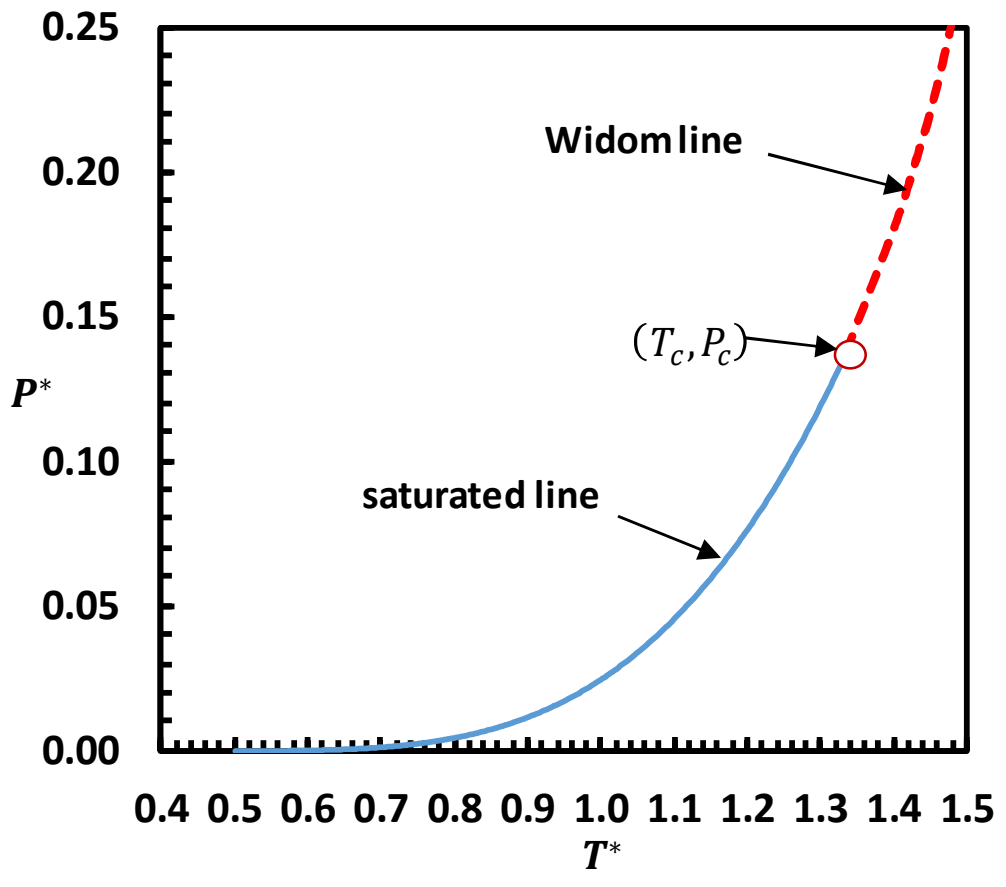

**Figure 10**. Phase diagram in the temperature-pressure $(T^*, P^*)$ plane. The saturated curve (solid line) is produced with the data from Ref.[25].

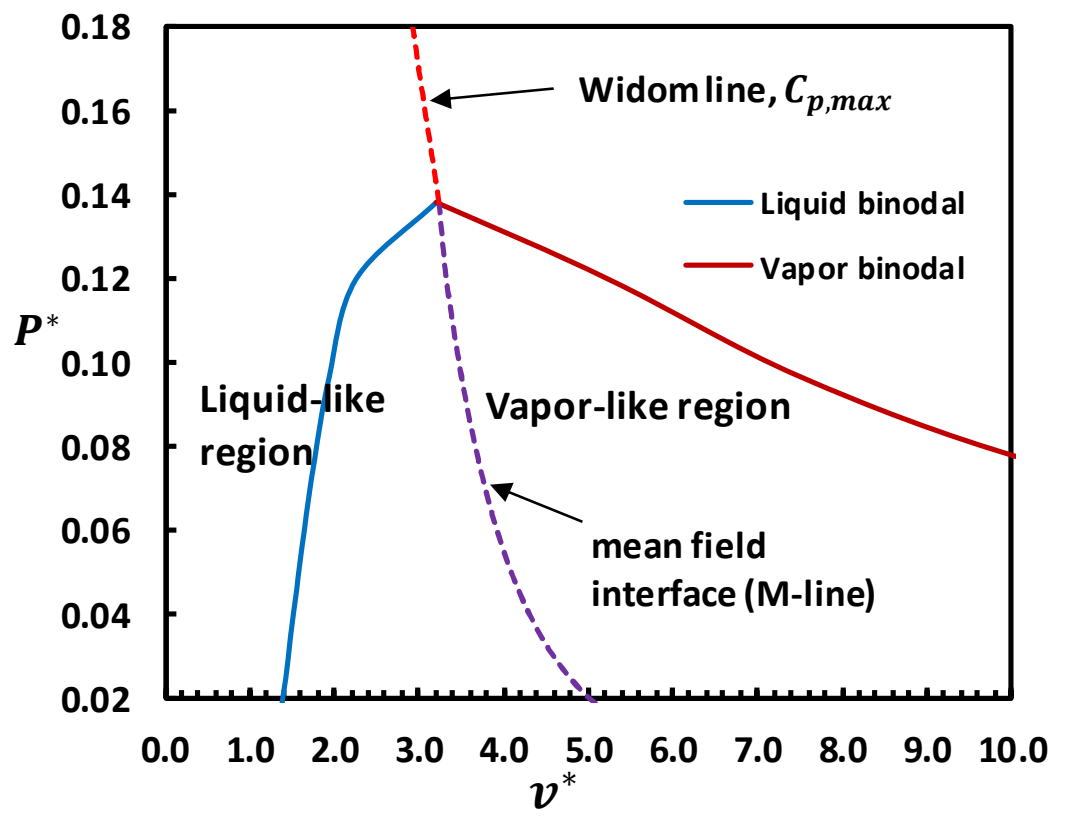

**Figure 11**. Phase diagram in the volume-pressure $(v^*, P^*)$ plane.

After the M-line (the mean-field interface) and the Widom line are both determined, we can produce the phase diagrams in all three (temperature-pressure-density) planes for the entire phase space. In this work the "entire phase space" refers to the subcritical (coexistence) region and the supercritical area that the Widom line divides into gas-like and liquid-like sub-regions.

Figure 9 - Figure 11 illustrate the phase diagrams in all three planes. An immediate observation is that in all the planes the mean-field dividing interface is the natural continuation of the Widom line into the two-phase coexistence region, or vice versa. This resolves a long-standing inconsistency: the Widom line is a smooth continuation of the saturated pressure in the pressure-temperature plane while in the density-temperature and density-pressure planes it connects to two branches at the critical point. This may be better stated with some mathematical arguments.

If a function or property, $f$, (pressure or density) and it's derivative $f'$ are equal , respectively, as approaching to the critical point from opposite directions: $f(T_c^+) = f(T_c^-)$ and $f'(T_c^+) = f'(T_c^-)$ we define this continuation as a $C^1$ level. If the values of the property are equal, $f(T_c^+) = f(T_c^-)$ while it's derivatives are not, $f'(T_c^+) \neq f'(T_c^-)$, the continuation is at $C^0$ level. With the conventional view, we face an inconsistency: in the pressure-temperature plane, the continuation is at $C^1$ level, while in the temperature-density and pressure-density planes, it is at $C^0$ level. In contrast, the continuation of the M-line (the mean-field interface) with the Widom line is at $C^1$ level in all three planes.

For the van der Waals fluid, an analytical expression for the Widom line can be obtained [19] and we can applied the above definitions strictly in all the temperature-pressure-density planes [28]. Here for the LJ fluid numerical results shown in Figure 9 – Figure 11 server the purpose. The slight inconsistency of the derivative, $dT^*/d\rho^*$, (Figure 9) is caused by the discrepancies between the saturated densities from two sources (see Appendix B, Figure B3). The main conclusion is that the entire phase space can be divided into gas-like and liquid-like regions by the M-line united with the Widom line. The M-line in the coexistence region is physically coherent with the Widom line in the supercritical region and hence the "single" line composed of the two acts as the unique demarcation line for the entire phase space.

In the subcritical region "static" nanoclusters exist in unstable states ($C_P < 0$) while in the supercritical region dynamic nanoclusters exist in stable states ($C_P > 0$). Here the term "static" does not exclude the size changes of the clusters in the interfacial region. The overall behavior of the isobaric heat capacity in the entire phase space can be summarized with the following state (denoted as "{ }") flow: $\{T_r < 0.65, C_P > 0\}$ => $\{0.65 < T_r < 1.0, C_P < 0\}$ => $\{T_r = 1.0, C_P \to \infty\}$ => $\{T_r > 1.0, C_P > 0\}$.

Finally, we briefly investigate the adsorption with the mean-field dividing interface. The local excess (or deficit) density is plotted in Figure 12. This figure should be read together with Figure 1. Firstly, on the vapor side, $\rho(z) - \rho_v > 0$ while on the liquid side, $\rho(z) - \rho_L < 0$. Secondly, at low temperature the adsorbed layer on the vapor side is much thinner than that on the liquid side. Finally, as the temperature approaches to the critical point, the difference of the two layers vanishes. All the above observations are consistent with experimental observations. In contrast, with the algebraic mean interface, Eq.(23), the amount of excess on the vapor side is the same as that of deficit on the liquid side at a given temperature.

the vapor and liquid layers are always the same in magnitudes (opposite signs) at a given temperature.

Figure 13 illustrates the calculation results for the total adsorption on the mean-field interface at the vapor side from Eq.(5). The unit is number of molecules per $\sigma^2$. These values are in consistent with the adsorption data reported in literature [32,33]. Since the mean-field interface does not satisfy Eq.(7) the adsorption defined by Eq.(5) and Eq.(6) can seen as referencing to that defined by Eq.(7). Extensive analysis and discussion of the subject are beyond the scope of this paper.

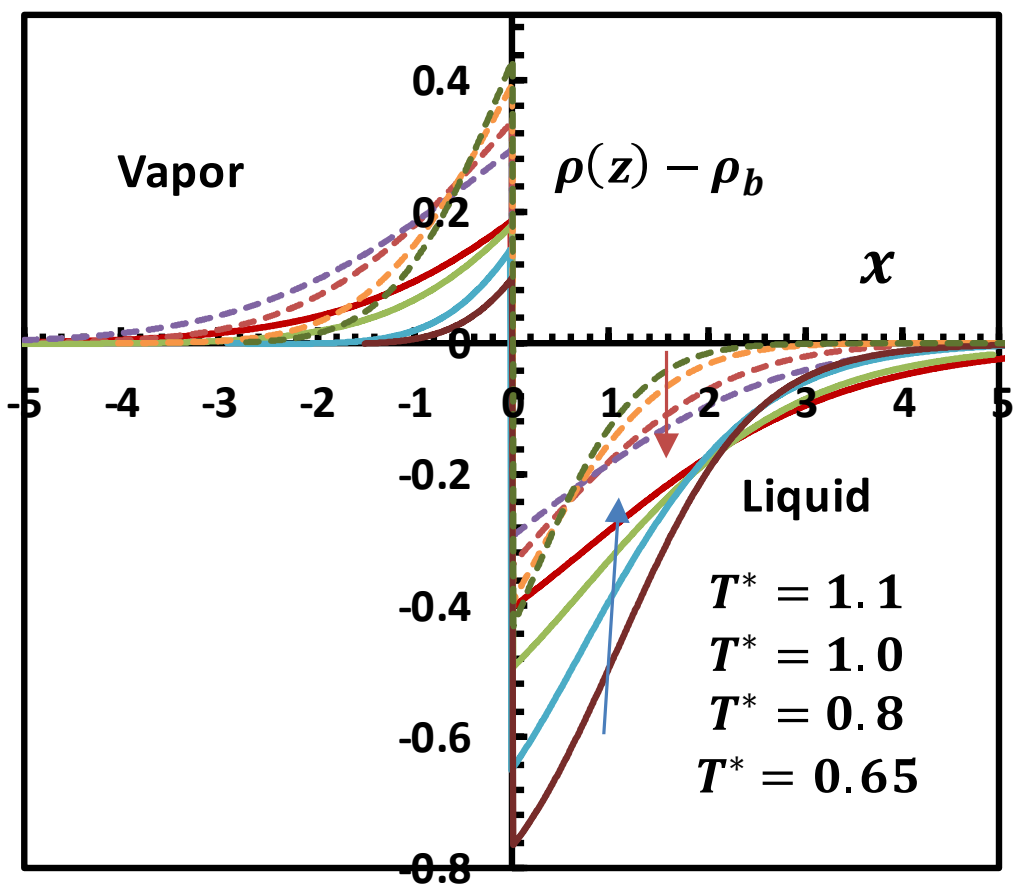

**Figure 12**. A comparison of the local density differences between those from Eq.(26) (solid lines) and those from Eq.(23) (dashed lines).

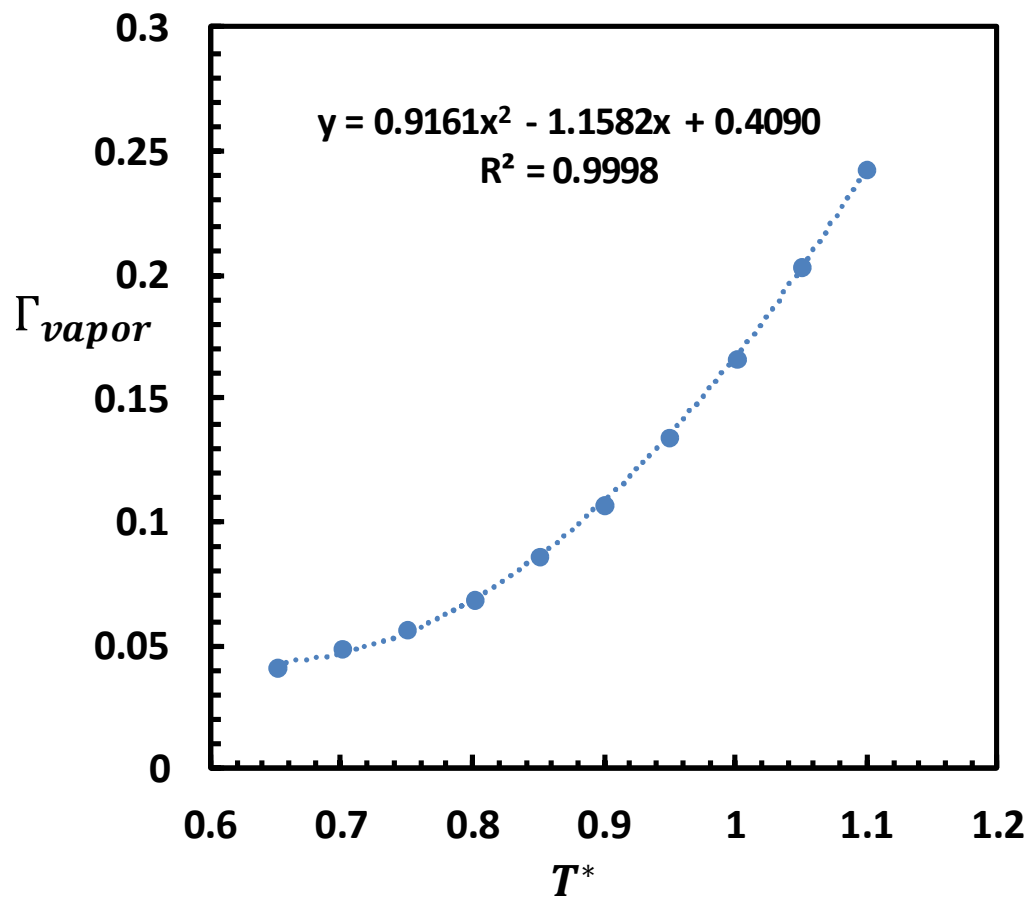

**Figure 13**. Total adsorption at vapor side calculated by Eq.(5) and Eq.(26).

## Conclusions and Discussions

This paper presents new findings for the vapor-liquid interface system by using the density gradient theory combined with the mean-field theory. A highly accurate expression for the density profile is proposed in which the mean-field Maxwell crossover, or the M-line, is employed as the dividing interface for the interfacial region. This new expression deals with the boundaries at vapor side and at liquid side separately to reflect the fact that the decay length of the vapor boundary is less than that of the liquid boundary [6]. The high accuracy of this new model guarantees that the intrinsic properties given by the density gradient theory are physically reliable, as confirmed by the excellent predictions of the pressure difference, $p_N(z) - p_T(z)$.

Using the density gradient theory and a mean-field LJ EoS [25], we are able to obtain analytical expressions for various properties for the LJ fluid in the interfacial region, such as the intrinsic Helmholtz free energy, the intrinsic Gibbs free energy and heat capacities, as functions of local position. It is shown that at the mean-field interface the intrinsic Helmholtz free energy exhibits a maximum value. The intrinsic Gibbs free energy peaks at $z - z_0 \approx 1$, corresponding to the same behavior of the pressure difference, $p_N(z) - p_T(z)$ at the liquid side. For the LJ fluid, it is found that the intrinsic isobaric heat capacity exhibits local maximum at the mean-field interface as a result of the heterogeneous contribution. All the phase diagrams show that the Widom line is smoothly (at $C^1$ level) connected to the M-line in all the planes. These observations firmly suggest that the M-line is the natural extension of the Widom line in the supercritical region into the coexistence region, or vice versa. Consequently, the single line composed of the M-line and the Widom line inherently divides the entire phase space into liquid-like and gas-like regions. In other words, the liquid-like and gas-like features in the supercritical region are inherited from the coexistence region via the mean-field dividing interface.

Another interesting feature is the sign of the heat capacity. In a stable fluid, heat capacity at constant volume is related to the mean-square fluctuation of internal energy, $C_v = \overline{\Delta E^2}/k_B T^2$, and is thus always positive, and $C_P > C_v$. In the interfacial region, as the system is composed of heterogeneous nanoscale clusters, the heat capacities show negative values [29-31]. Some argue [30] that "negative heat capacity in nanoclusters is an artifact of applying equilibrium thermodynamic formalism to a cluster…". On the other hand, theoretical analysis [29] and experiment [31] reported negative heat capacities in some nanocluster systems. The results obtained from the density gradient theory and the mean-field theory for the LJ fluid unveil a more complicated picture.

First of all, the behavior of the isobaric heat capacity contradicts with the claim of Ref.[30] since negative values are obtained with the intrinsic heat capacity where the heterogeneous contribution is added to the mean-field formalism. In addition, the homogeneous mean-field contribution alone can be either positive or negative in the interfacial region (Figure 7c). Moreover, in the interfacial region, $C_v > C_P$. There is a "transition" temperature, $T^* \approx 0.85$ ($T_r \approx 0.65$), above which $C_P < 0$ and below which $C_P > 0$. Nevertheless, the feature of the local maximum of the isobaric heat capacity, $C_P$, is reserved in the entire phase space.

Now we can address the second-order phase transition from a differently perspective along the M-line and the Widom line. At a subcritical region ($1.0 \geq T_r \geq 0.65$ for the LJ fluid), the unstable nanoscale system exhibits negative intrinsic heat capacity. At the critical point, the heat capacity diverges (positively) and it becomes finite positive along the Widom line. Therefore, the critical point can be seen as the transition point where an unstable heterogeneous system composed of "static" nanoclusters becomes a stable homogeneous system composed of dynamic nanoclusters.

This work also breathes some new life into the classic mean-field theory subjected to the Maxwell construction for the VLE calculations. All the information required for completing the phase diagrams in the entire phase space are embedded in the theory. Three solutions (for a vast majority of EoS) have their respective roles or significances: two as the saturated volumes of vapor and liquid phases, respectively, which are related to the first-order transition and the third one as the dividing interface, which is related to the second-order transition. This feature is even better demonstrated with a cubic EoS based on the well-known van der Waals EoS [28,34].

Finally, the new density profile model provides an alternative avenue for prediction of the adsorption on the vapor-liquid interface. Primitive calculations show some promising results. Further investigations need to be done including comparing the predicted results with experimental data and direct computer simulations.

## Acknowledgements

The author is grateful to Dr. Langenbach and Dr. Stephan for helpful discussions and providing their research articles. Special thanks to Dr. Mejía for helpful discussions and kindly providing unpublished data of the density profile and the pressure differences, $p_N(z) - p_T(z)$ for the LJ fluid in the temperature range $0.65 \leq T^* \leq 1.1$.

## Appendix A. Density profile equation

In this appendix, we derive a new density profile expression. Two routes can be taken to derive the governing differential equation for the density profile [6]: from the Kirkwood-Buff theory or from the statistical mechanics method (direct correlation function). Here we adopt the Kirkwood-Buff approach. In the original solution to the governing differential equation, the approximation made in Ref.[6] leads a simple exponential function, which is apparently not accurate enough and inconsistent with those employed in the literature [2,23], such as Eq.(23). In this work we revisit the solution by using a more appropriate approximation.

In the interfacial region two coordinate systems are used: (1) for the position of a planar surface a unique coordinate normal to the surface, $z$, is used; (2) for a particle/molecule in the interfacial region, the radial coordinate system, $\bar{r}$ $(r, \theta, \varphi)$, is employed. Considering two particles on different surfaces in system (1) at positions, $z_1$ and $z_2$, respectively, their distance is denoted as $z_{12}$ and their positions in system (2) are denoted as $\bar{r}_1$ and $\bar{r}_1$, respectively, with a distance, $r_{12}$.

According to the Kirkwood-Buff theory, for a planar surface the normal component (perpendicular to the surface) of the stress tensor at $z_1$ is given by [6]:

$$\sigma_N(z_1) = \beta^{-1}\rho(z_1) - \frac{1}{2}\int \frac{z_{12}}{r_{12}}\rho^{(2)}(\bar{r}_1, \bar{r}_1)\frac{du(r_{12})}{dr_{12}}d\bar{r}_{12} (A1)$$

where $\beta = 1/k_B T$, $u(r_{12})$ is the inter-molecular potential, $\rho^{(2)}(\bar{r}_1, \bar{r}_1)$, the pairwise (particle) distribution function. At hydrostatic equilibrium, $\sigma_N(z_1) = constant$, therefore:

$$\beta^{-1}\frac{d\rho(z_1)}{dz_1} = \int \frac{z_{12}}{r_{12}}\rho(z_2)g(\rho_b; r_{12})\frac{du(r_{12})}{dr_{12}}d\bar{r}_{12} \qquad (A2)$$

where the relation $\rho^{(2)}(\bar{r}_1, \bar{r}_1) = \rho(z_2)g(\rho_b; r_{12})$ has been introduced with $g(\rho_b; r_{12})$ being the radial distribution function and $\rho_b$, the density of the bulk phase, which is either $\rho_v$ for the vapor phase, or $\rho_L$ for the liquid phase. The integration of equation (A2) yields the density profile [6]:

$$\beta^{-1}ln\frac{\rho(z)}{\rho_b}=$$

$$\int_0^z\left[\iint\frac{z_{12}}{r_{12}}\rho(z_1+z_{12})g(\rho_b;r_{12})\frac{du(r_{12})}{dr_{12}}d\bar{r}_{12}\right]dz_1 \quad (A3)$$

By assuming that $\rho(z)$ varies slowly over the range of inter-molecular force, $du(r)/dr$, then $\rho(z_1+z_{12})$ can be expanded around $\rho(z_1)$. Next by exchanging the integration over $d\bar{r}_{12}$ in the radial coordinate system into that in the cylindrical coordinate system, $(r,z,\varphi)$, one gets the governing equation:

$$ln\frac{\rho(z)}{\rho_b}=\alpha_0[\rho(z)-\rho_b]+\alpha_2\frac{d^2\rho(z)}{dz^2}+\alpha_4\frac{d^4\rho(z)}{dz^4}+\cdots \quad (A4)$$

where the odd terms are vanished. Therefore, if we truncate right side of Eq.(A4) at the 2nd order while omitting the 4th and higher order terms, the resulting equation is effectively exact up to the 3rd order. The coefficients in eq.(A4) are defined as [6]:

$$\alpha_0=\frac{4\pi\beta}{3}\int_0^\infty r^3g(\rho_b;r)\frac{du(r)}{dr} \quad (A5a)$$

$$\alpha_2=\frac{2\pi\beta}{15}\int_0^\infty r^5g(\rho_b;r)\frac{du(r)}{dr} \quad (A5b)$$

where $\alpha_0>0$, $\alpha_2>0$. This feature is of critical importance for solving Eq.(A4). Now we apply Eq.(A4) to the vapor phase by dropping the 4th order and higher order terms:

$$\alpha_2\frac{d^2\rho_v(z)}{dz^2}+\alpha_0(\rho_v(z)-\rho_v)-ln\frac{\rho_v(z)}{\rho_v}=0 \quad (A6)$$

Multiplying Eq.(A6) by $\frac{d\rho_v(z)}{dz}$ and integration yield:

$$\alpha_2\left(\frac{d\rho_v(z)}{dz}\right)^2+\alpha_0[\rho_v^2(z)-\rho_v^2]+2(1-\alpha_0\rho_v)[\rho_v(z)-\rho_v]=2\rho_v(z)ln\frac{\rho_v(z)}{\rho_v} \quad (A7)$$

We define a new variable:

$$\bar{\rho}=\rho_v(z)-\rho_v>0 \quad (A8)$$

By considering the fact that the above differential equation is exact up to the 3rd order, the expansion of the logarithm term ought to be written as:

$$ln\frac{\rho_v(z)}{\rho_v}=ln\left(\frac{\bar{\rho}}{\rho_v}+1\right)\approx\frac{\bar{\rho}}{\rho_v}-\frac{1}{2}\frac{\bar{\rho}^2}{\rho_v^2}+\frac{1}{3}\frac{\bar{\rho}^3}{\rho_v^3} \quad (A9)$$

By inserting Eq.(A9) into Eq.(A7), after some rearrangements, we obtain:

$$\left(\frac{d\bar{\rho}}{dz}\right)^2=b\bar{\rho}^2-a\bar{\rho}^3 \quad (A10)$$

where

$$a=\frac{1}{3\alpha_2}\frac{1}{\rho_v^2}>0 \quad (A11a)$$

$$b=\frac{1}{\alpha_2}\left(\frac{1}{\rho_v}-\alpha_0\right)>0 \quad (A11b)$$

The last arguments, Eq.(A11a) and Eq.(A11b), need some elucidating since they are critical for obtaining the final solution. From the definitions, Eq.(A5a), Eq.(A5b), we can estimate the values of $\alpha_0$ and $\alpha_2$ when a radial distribution function is known. For the LJ fluid analytical expressions of radial distribution function (RDF) can be found in the literature. By using a simple analytical expression [35], some computations have been carried out. At T*=0.65, the value of $\alpha_0$ is in order of ~10, while $1/\rho_v\sim10^3$, and at T*=1.1, $\alpha_0$ is in order of $10^0$, while $1/\rho_v\approx18$. In addition, in the entire temperature range, the value of $\alpha_2$ is between 1 and 4. Hence Eq.(S11) holds for the entire temperature range considered here. The calculations also show that $b$ and $a\bar{\rho}$ are in the same order of magnitude and thus ignoring the term $a\bar{\rho}^3$ in Eq.(A10) (as did in Ref.[6]) is not acceptable.

For calculation of the density profile, an accurate RDF is required, and the solutions could only be numerical since only numerical integration of Eq.(A5) is possible. Moreover, solving Eq.(A4) has to be done for the vapor branch and liquid branch, respectively. Since the goal of this work is to derive a single analytical function for the density profile we will not perform any numerical calculations.

For integrating Eq.(A10), a new variable is introduced: $u=\sqrt{b-a\bar{\rho}}$ , and by noticing that $a\bar{\rho}=b-u^2>0$, Eq.(A10) can be written as:

$$-\frac{1}{2}\frac{d\bar{\rho}}{\bar{\rho}\sqrt{b-a\bar{\rho}}}=\frac{du}{b-u^2}=-\frac{1}{2}dz \quad (A12)$$

On integration [36], we have:

$$\int \frac{du}{b-u^2} = \frac{1}{\sqrt{b}} atanh\left(\frac{u}{\sqrt{b}}\right) + C = \frac{1}{2}(z-z_0) \quad (A13)$$

In the current model, we define $z_0$ in such a manner that the region $-\infty < (z-z_0) < 0$ is for the vapor side, and $0 < (z-z_0) < \infty$ for the liquid side. Therefore in the vapor side, $z < z_0$, and in Eq.(A13), the sign has been changed so that the same shifted variable $z-z_0$ can be applied to both liquid and vapor sides at the same time. According to the properties of the hyperbolic functions [36], Eq.(A13) can be rewritten as:

$$u = \frac{\tanh(-C\sqrt{b}) + thanh\left[\frac{\sqrt{b}}{2}(z-z_0)\right]}{1+\tanh(-C\sqrt{b})\, thanh\left[\frac{\sqrt{b}}{2}(z-z_0)\right]} \quad (A14)$$

The conditions for obtaining Eq.(A14) are $b > 0$ and $b-u^2 > 0$ [36] and the definitions of $b$ and $u$ guarantee that both conditions are met as discussed above. Since$|\tanh(x)| < 1$, we can neglect all the terms with $\tanh^2(x)$ ($\bar{\rho} = \rho_v(z) - \rho_v = b/a - u^2/a$) and finally after some straightforward algebra we have:

$$\rho(z) = A + \frac{B\tanh\left(\frac{z-z_0}{D}\right)}{1+C\tanh\left(\frac{z-z_0}{D}\right)} \quad (A15)$$

where $A$, $B$ and $C$ are (bulk) density-dependent, and $D = 2/\sqrt{b}$, which is a quantity related to the thickness of the interfacial region. Eq.(A15) is derived for the vapor side of the interface ($\rho(z)$ should be read as $\rho_v(z)$ ). For the liquid side a similar treatment can be done starting from Eq.(A4). However, for obtaining a single density profile equation, we empirically "merge" the vapor side and liquid side with the following form:

$$\rho(z) = A + \frac{B\tanh\left(\frac{z-z_0}{D_v}\right)}{1+C\tanh\left(\frac{z-z_0}{D_L}\right)} \quad (A16)$$

where $D_v$ is related to the thickness of the vapor layer, and $D_L$ to that of the liquid layer. Hence, $D_v + D_L$ is related to the total thickness of the interface area, corresponding to the parameter $D$ in the conventional (classic) model, Eq.(23). The three parameters, A, B and C can be determined by the conditions discussed in the main text. From Eq.(A16), the derivatives can be easily obtained:

$$\frac{d\rho(z)}{dz} = \frac{B\,\mathrm{sech}^2\frac{z-z_0}{D_v}}{1+C\tanh\frac{z-z_0}{D_L}}\left[\frac{1}{D_v} - \frac{1}{D_L}\frac{C\tanh\frac{z-z_0}{D_v}}{1+C\tanh\frac{z-z_0}{D_L}}\right] (A17)$$

$$\frac{d^2\rho(z)}{dz^2} = -2\frac{d\rho(z)}{dz}\left(\frac{1}{D_v}\tanh\frac{z-z_0}{D_v} + \frac{1}{D_L}\frac{C\,\mathrm{sech}^2\frac{z-z_0}{D_v}}{1+C\tanh\frac{z-z_0}{D_L}}\right) (A18)$$

where:

$$B = 2(\rho_L - \rho_M)\left(\frac{\rho_M - \rho_v}{\rho_L - \rho_v}\right) \quad (A19)$$

$$C = \frac{2\rho_M - \rho_v - \rho_L}{\rho_L - \rho_v} \quad (A20)$$

Eq.(A17)-(A20) are used for calculations of density gradients in Eq.(37) and Eq.(38).

## Appendix B. The LJ EoS, VLE and heat capacity

There are two equivalent ways to determine the equilibrium (saturated) properties from an EoS. First, the Maxwell construction in the pressure-volume plane, which is also known as the equal-area rule, as shown in Figure B1. The M-line is determined by:

$$\int_{v_L}^{v_M} (P^{*s} - P)\, dv = \int_{v_M}^{v_G} (P - P^{*s})\, dv \quad (B1)$$

or area ABC=area CDE. The second way is illustrated by the inset of Figure B1, namely, in the pressure-chemical potential plane:

$$\mu^{*L} = \mu^{*v} \quad (B2)$$

The condition, $P^*(v_L) = P^*(v_G) = P^{*s}$, is applied in both methods. By the way, as shown in Figure B1 (inset), a temperature (only) -dependent term in Eq.(B2) and Eq.(35) will be cancelled.

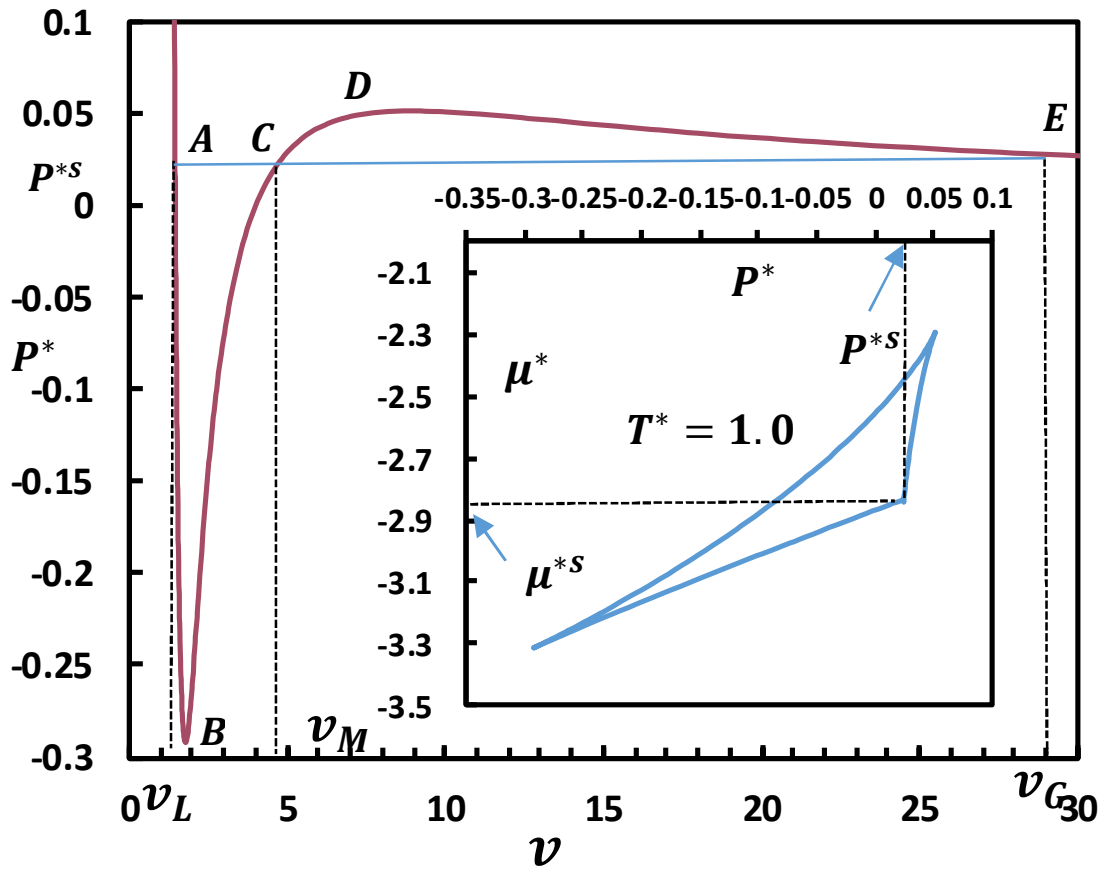

**Figure B1**. Determinations of the equilibrium properties and the Maxwell crossover, $v_M$, by the Maxwell construction in the $P \sim v$ plane. Inset: determination of equilibrium chemical potential and pressure in the $\mu \sim P$ plane.

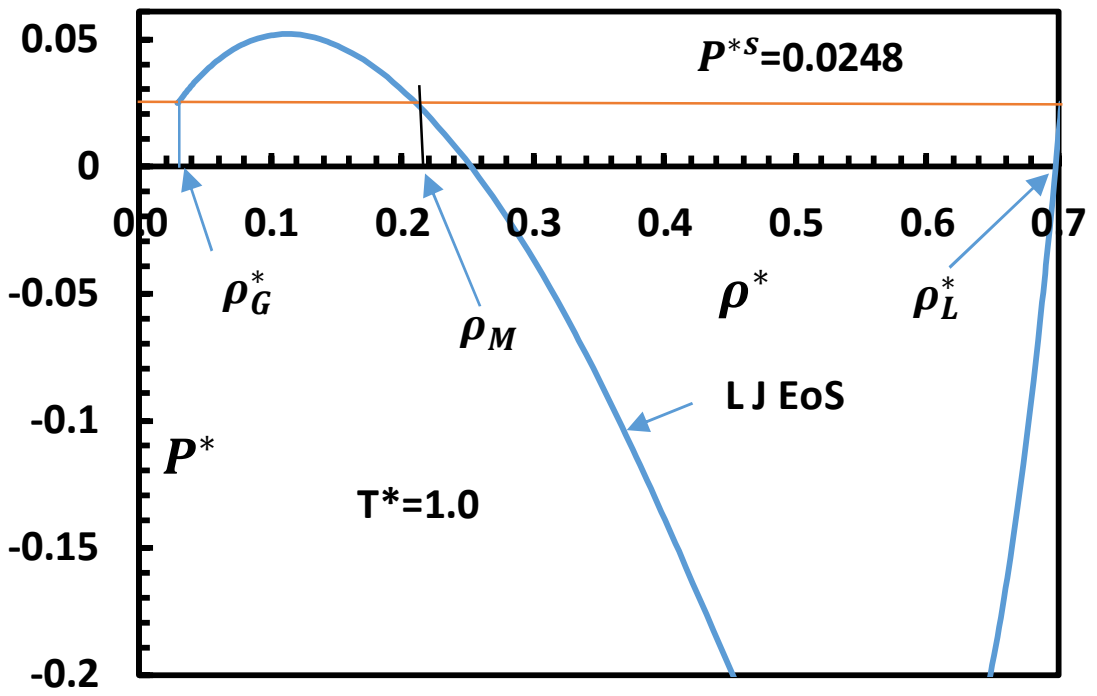

**Figure B2**. Determination of $\rho_M$ from a given saturated pressure in the pressure-density plane at a given temperature.

If the saturated pressure is known at a given temperature, one can determine the M-line directly in the pressure-density plane by using the method illustrated in Figure B2. The intersection of the vapor pressure with the density curve generated by an EoS gives $\rho_M$. In this work the saturated densities and pressure are from Ref [25], and the M-line is produced accordingly.

**Table B1** critical constants

| Ref | $T_c^*$ | $P_c^*$ | $\rho_c^*$ |
|---|---|---|---|
| [25] | 1.335 | 0.138 | 0.310 |
| [8,24] | 1.312 | 0.130 | 0.300 |

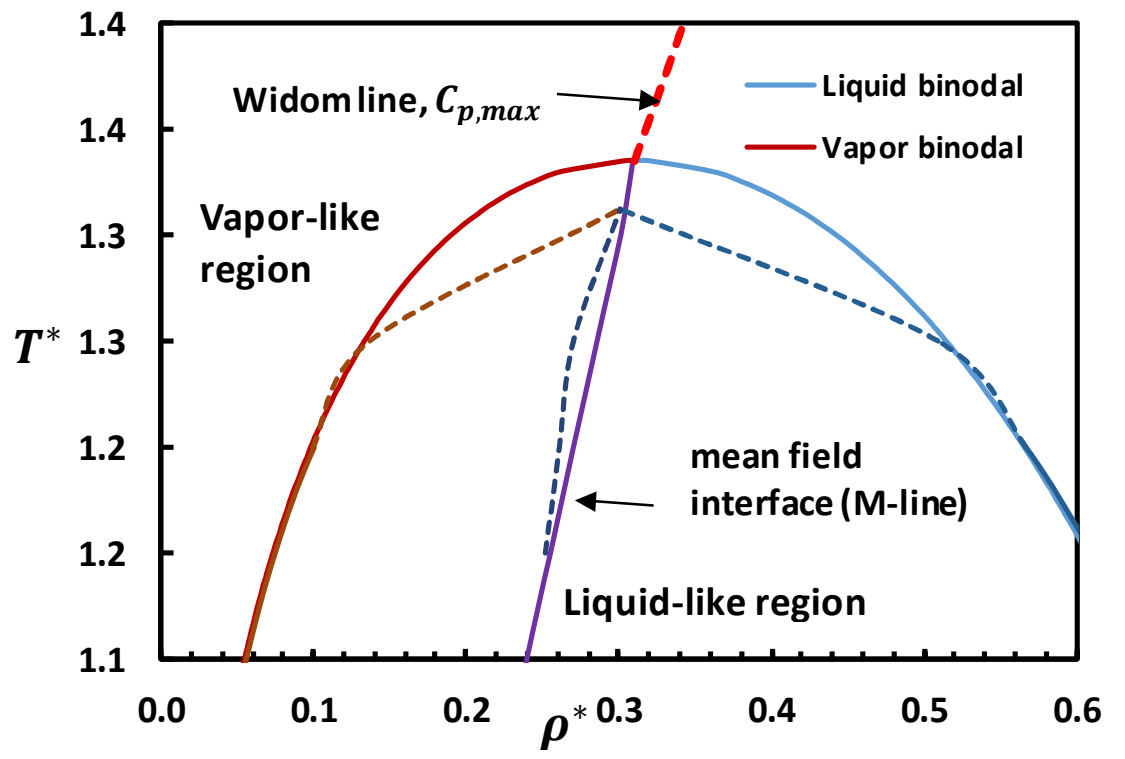

Figure B3. A comparison of saturated densities in the high temperature region. Solid M-line is calculated with saturated properties from Ref [25]; Dashed M-line with saturated properties from Ref.[8.24].

The data for the saturated fluids from two sources (Ref (8,24) and Ref [25]) are not entirely consistent with each other. Table B1 lists the critical constants. Figure B3 illustrates the saturated densities in high temperature region, along with the M-lines generated from different saturated densities. This discrepancy contributes the most to the deviations of the maxima of the isobaric heat capacity from the mean-field interface (Figure 7b).

The LJ EoS used here is from Ref [25]. Since we only consider a pure fluid the reference term is simplified to the Carnahan-Starling EoS. The residual free energy is given by:

$$\tilde{a}_{res} = \tilde{a}_{ref} + \tilde{a}_{pert} \qquad (B3)$$

$$\tilde{a}_{ref} = \tilde{a}_{hs} = \frac{4\eta - 3\eta^2}{(1-\eta)^2} \qquad (B4)$$

$$\tilde{a}_{pert} = \frac{a_{pert}}{T^*} = \frac{\tilde{a}_1}{T^*} + \frac{\tilde{a}_2}{T^{*2}} \qquad (B5)$$

$$\tilde{a}_1 = -2\pi\rho^* \sum_{i=0}^{7} a_i \eta^i \qquad (B6)$$

$$\tilde{a}_2 = -\pi\rho^* \left[1 + \frac{8\eta - 2\eta^2}{(1-\eta)^4}\right]^{-1} \sum_{i=0}^{7} b_i \eta^i \qquad (B7)$$

where the effective packing fraction is defined as:

$$\eta = \frac{\pi}{6}\rho^* d^3 \qquad (B8)$$

where $d(T)$ is known as the effective diameter. Eq.(B4) is from the Carnahan-Starling EoS for the hard sphere fluid and the effective diameter is given by:

$$d(T) = 1 - c_1 exp\left(-\frac{c_2}{T^*}\right) \qquad (B9)$$

The values of the constants, $a_i$, $b_j$ and $c_k$ are listed in Table B2.

**Table B2**. Coefficients of the LJ EoS [25]

| | | | | | |
|---|---|---|---|---|---|
| $a_0$ | 0.858981 | $b_0$ | 0.640858 | $c_1$ | 0.114289 |
| $a_1$ | 0.634967 | $b_1$ | 3.295024 | $c_2$ | 2.913319 |
| $a_2$ | 6.397088 | $b_2$ | -2.07167 | | |
| $a_3$ | -39.4901 | $b_3$ | -201.226 | | |
| $a_4$ | 109.8421 | $b_4$ | 1151.68 | | |
| $a_5$ | -173.332 | $b_5$ | -3855.85 | | |
| $a_6$ | 139.7734 | $b_6$ | 8261.241 | | |

For the homogeneous contributions to the heat capacities, we have [25]:

$$C_v = -\left(\tilde{a}_{20}^{id} + \tilde{a}_{20}^{res}\right) \qquad (B10)$$

$$C_P = C_v + \frac{(1 + \tilde{a}_{01}^{res} - \tilde{a}_{11}^{res})^2}{1 + 2\tilde{a}_{01}^{res} + \tilde{a}_{02}^{res}} \qquad (B11)$$

$$\tilde{a}_{nm}^{res} = \tau^n \rho^{*m} \frac{\partial^{n+m}(\tilde{a}^{res})}{\partial\tau^n \partial\rho^{*m}} \qquad (B12)$$

**Table B3** Coefficients of Eq.(B15) for $0.7 \leq T^* \leq 1.0$

| $k$ | $c_{4,k}$ | $c_{3,k}$ | $c_{2,k}$ | $c_{1,k}$ | $c_{0,k}$ |
|---|---|---|---|---|---|
| 1 | 0 | 0 | 0 | 0 | -8.6063 |
| 2 | 8.2965 | -45.995 | 51.256 | -16.746 | 31.852 |
| 3 | -135.539 | 321.783 | -251.605 | 64.538 | -37.385 |
| 4 | 119.841 | -255.94 | 183.597 | -43.125 | 14.011 |

For an interface system, the derivatives in Eq.(B12) should be applied to $\tilde{f}_2$, Eq.(42), as well. For instance:

$$\tilde{a}_{01}^{int} = \rho^* \frac{\partial\left(\tilde{f}_2 + \tilde{a}^{res}\right)}{\partial\rho^*} \qquad (B13)$$

Now we need to find the analytical temperature and density dependences of $\tilde{f}_2(\rho(z))$. The values of $\rho' = d\rho/dz$ are firstly calculated at each temperature by Eq.(A17), then dependences of $\tilde{f}_2(\rho(z))$, Eq.(42), on density and temperature are fitted with the following polynomial function:

$$\tilde{f}_2(\rho(z)) = \sum_{i=0}^{4} c_{4-i}\rho(z)^{4-i} \qquad (B14)$$

where the coefficients are temperature-dependent:

$$c_j = \sum_{k=0}^{3} c_{j,3-k}\tau^{(3-k)} \qquad (B15)$$

The coefficients of Eq.(B15), $c_{j,k}$ , are listed in Table B3. It is found that the accuracies of the generic correlation, Eq.(B14) and Eq.(B15), become low in the low ($T^* \leq 0.7$) temperature region and high $T^* \geq 1.0$) temperature region. This causes the less accuracies in heat capacity evaluations in those regions.

The derivatives required in calculation of the intrinsic heat capacities can be easily obtained from Eq.(B14) and Eq.(B15). For examples:

$$\frac{\partial^2 \tilde{f}_2}{\partial\rho^2} = \sum_{i=0}^{2}(4-i)(3-i)c_{4-i}\rho(z)^{2-i} \qquad (B16)$$

$$\frac{\partial^2 \tilde{f}_2}{\partial\rho\partial\tau} = \sum_{i=0}^{3}(4-i)\frac{dc_{4-i}}{d\tau}\rho(z)^{3-i} \qquad (B17)$$

The generic correlation with Eq.(B14) and Eq.(B15) is less accurate in low and high temperature regions. Figure 6 depicts the correlations for $\tilde{f}_2$ in the temperature range: $0.7 \leq T^* \leq 1.0$.

The calculation results for the isochoric heat capacity are illustrated in Figure B4. Figure B4a depicts the homogeneous contribution from the mean-field LJ EoS, Eq.(B10) and Eq.(B11), and all values are positive. The results for intrinsic isochoric heat capacity at different temperatures, Eq.(39), are plotted in Figure B4b. At temperature $T^* < 0.85$, $C_v > 0$, and $T^* > 0.85$, $C_v < 0$. By comparing Figure B4b with Figure 7b, it is seen that the sign of intrinsic isobaric heat capacity, $C_P$, is dominated by the sign of $C_v$. The local maxima of the intrinsic heat capacity are dominated by the heterogeneous contributions.

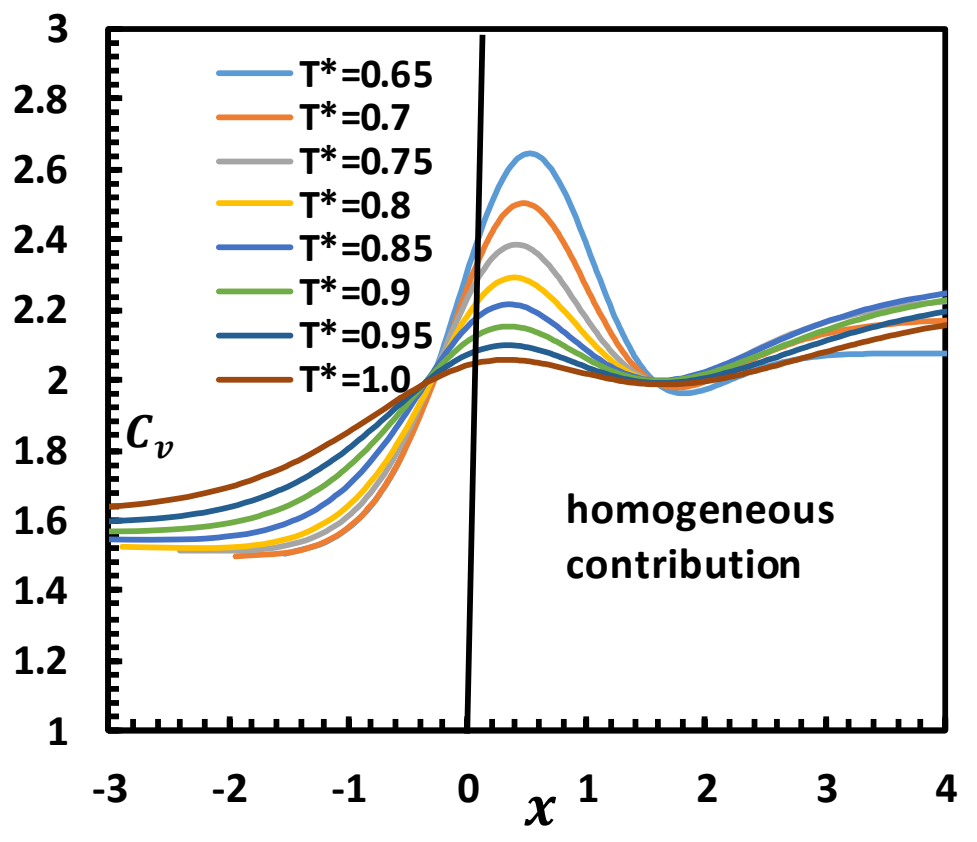

**Figure B4a**. Homogeneous (mean-field) contribution to the isochoric heat capacity.

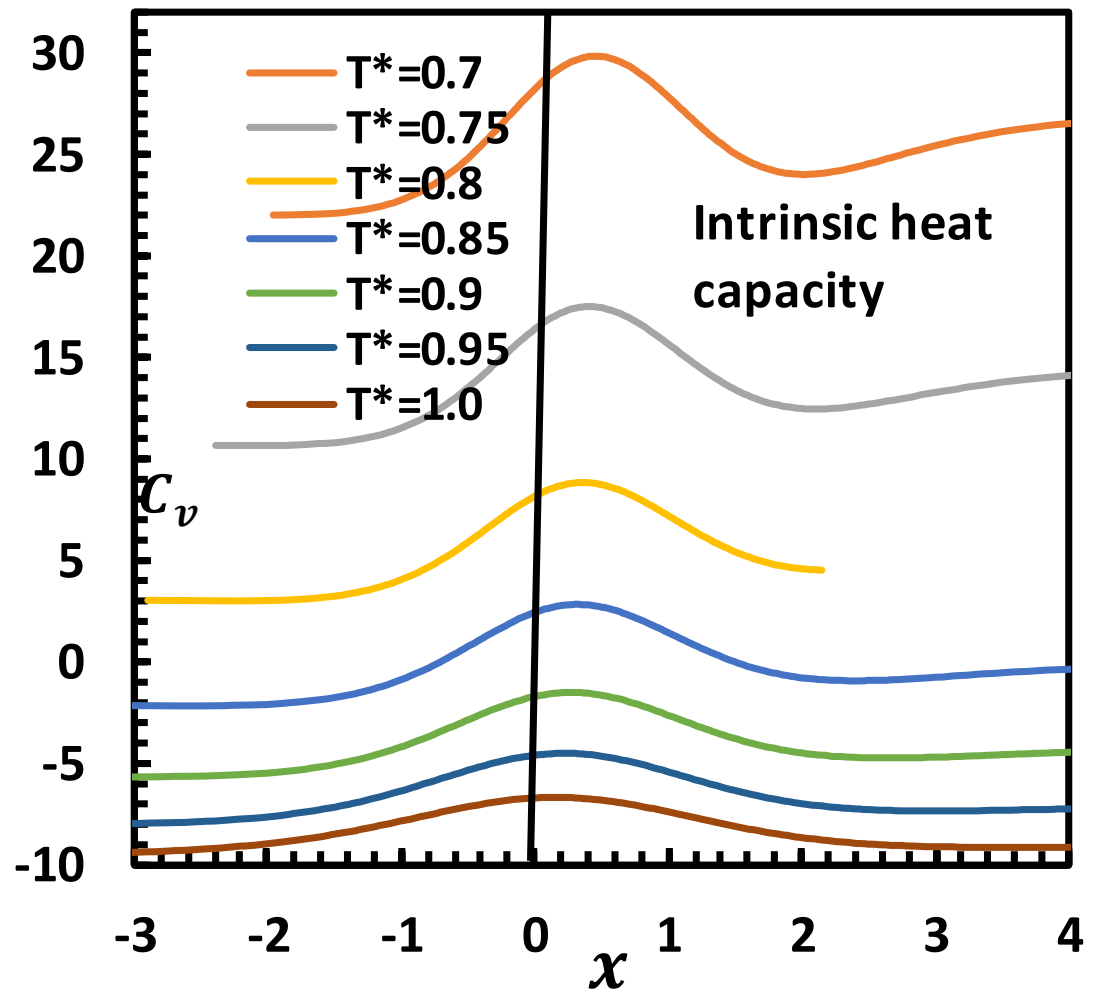

**Figure B4b**. Intrinsic isochoric heat capacity

**Figure B4**. Heat capacity at constant volume

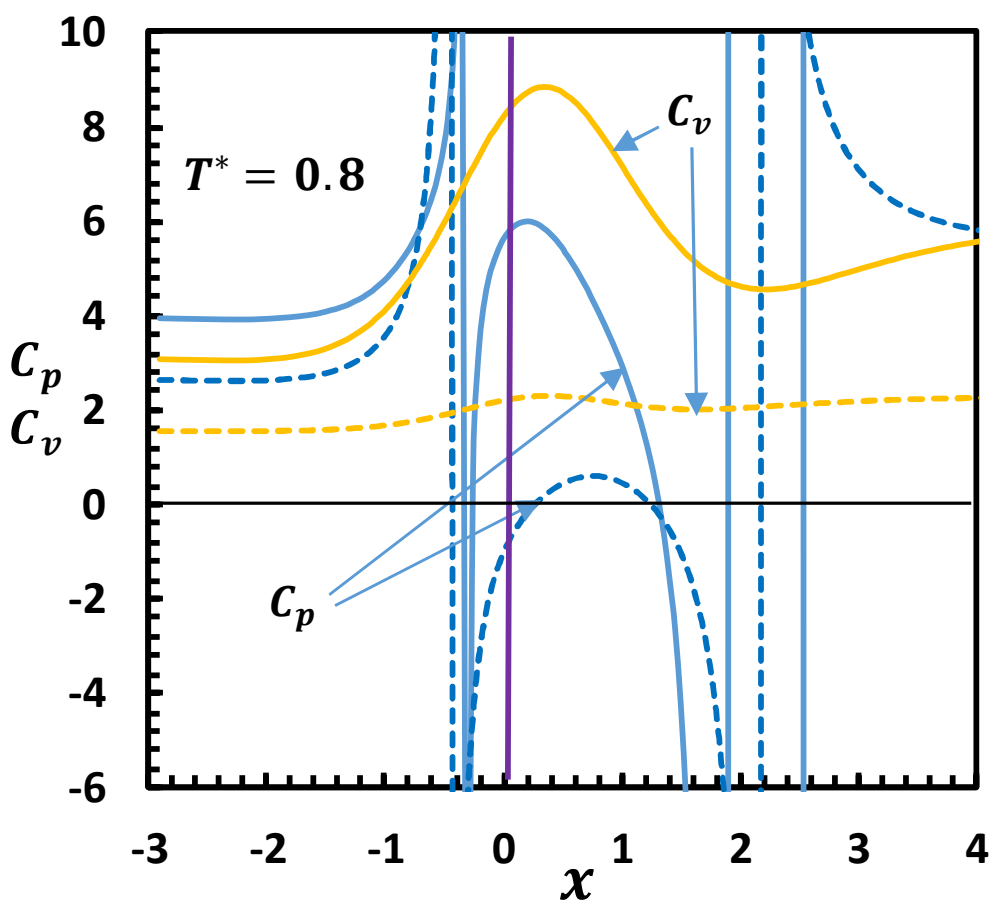

**Figure B5a**. Heat capacity at T*=0.8.

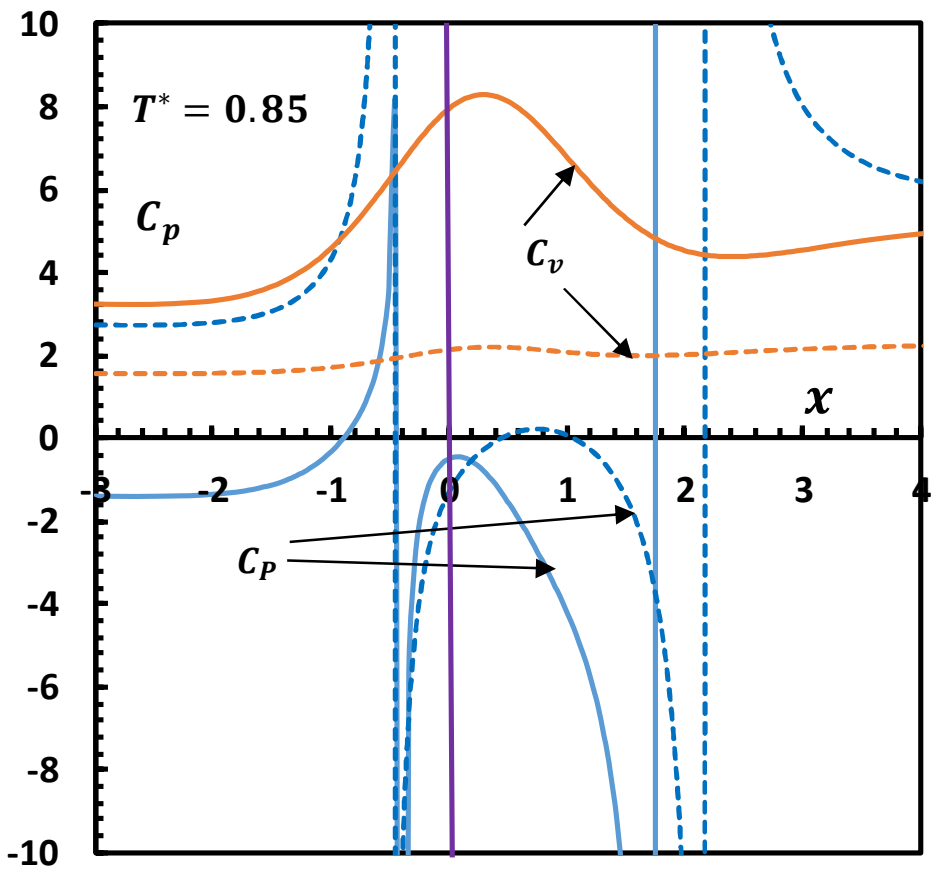

**Figure B5b**. Heat capacity at T*=0.85.

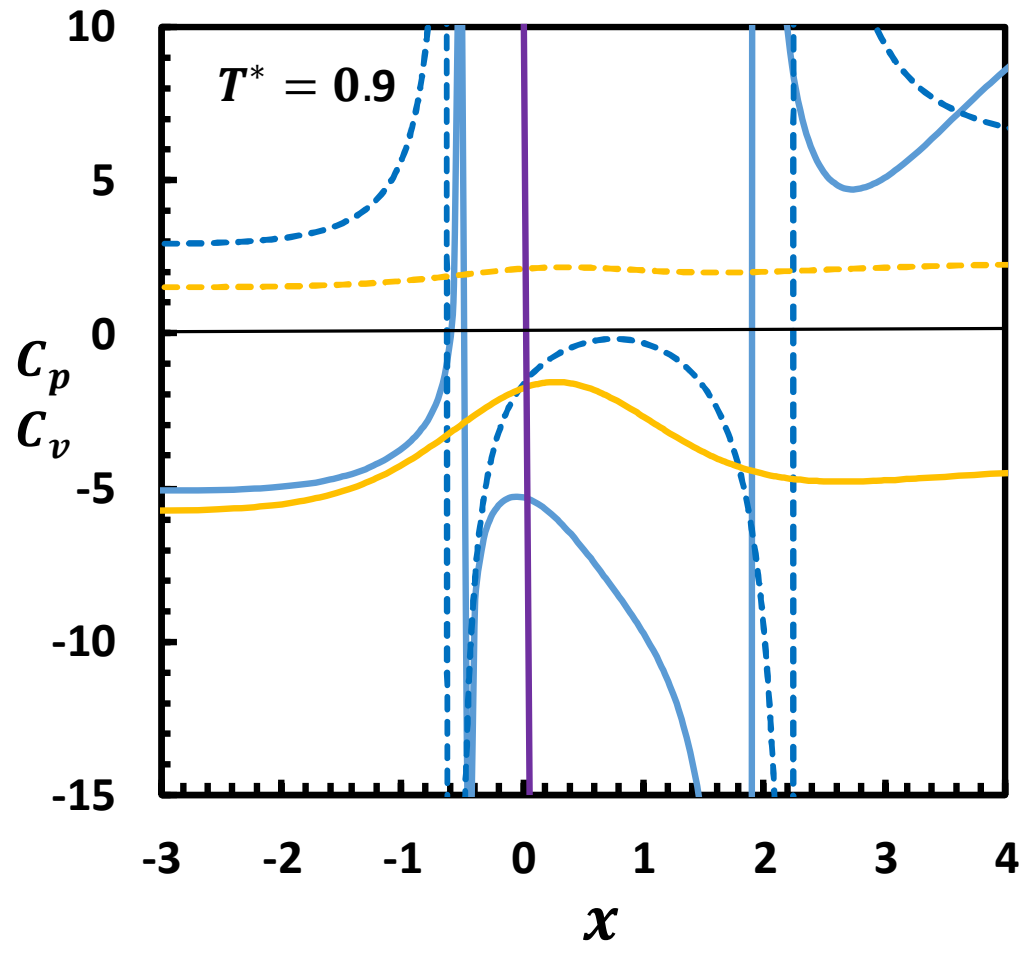

**Figure B5c**. Heat capacity at T*=0.9.

**Figure B5**. Comparison of heat capacities: solid lines: intrinsic heat capacities; dashed lines: homogeneous contributions.

Figure B5 illustrates comparisons of heat capacity at three temperatures. As $T^* > 0.85$, the isobaric heat capacity is negative while as At $T^* < 0.85$ it is positive.

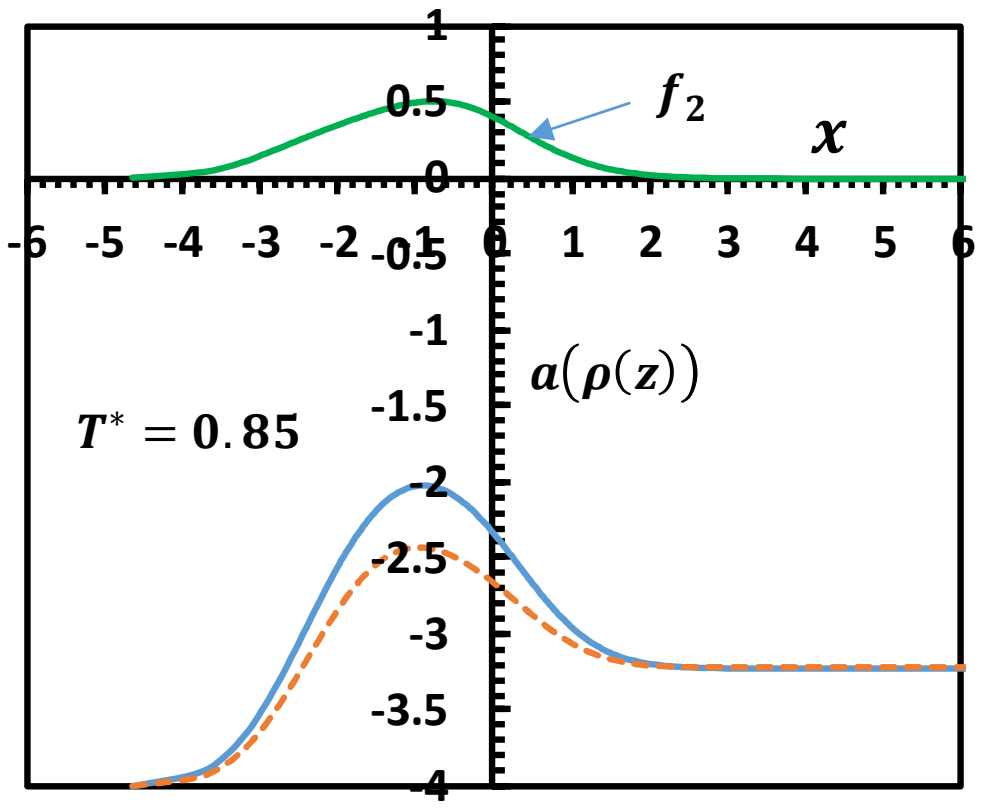

**Figure B6a**. Components of the Helmholtz free energy from the classic density profile. Solid blue line: the intrinsic free energy; dashed line: homogeneous contribution.

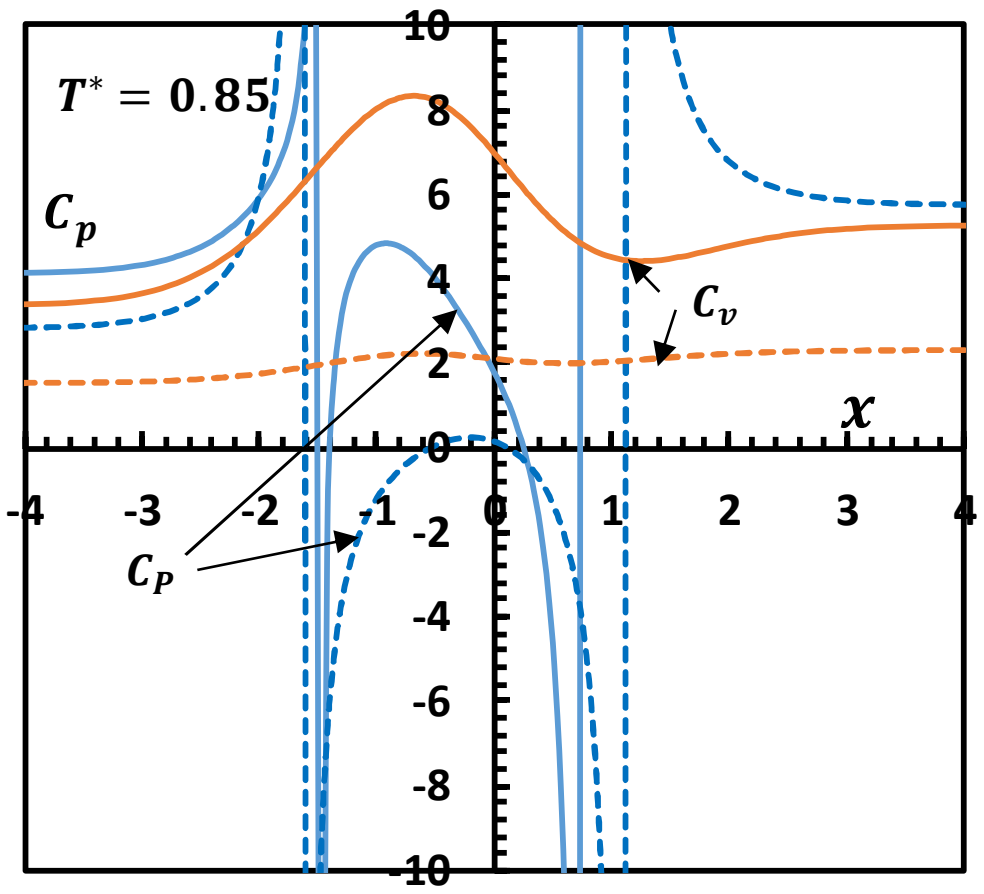

**Figure B6b**. Heat capacity calculated from the classic density profile. Solid lines: the intrinsic heat capacities; dashed lines: homogeneous contributions.

**Figure B6**. Free energy and heat capacity from the LJ EoS with the classic density profile, Eq.(23).

Finally, Figure B6 depicts the free energy and heat capacities calculated from the LJ EoS [25] with the classic density profile, Eq.(23). This figure shows that the local maxima appear at $x \approx -1.0$, not at $x = 0.0$ in contrast to the case with the mean-field density profile. The conclusion is that the location the maximum loci of the free energy and the isobaric heat capacity does not necessarily coincide with that of the dividing interface.